\documentclass[fleqn,usenatbib]{mnras}

\usepackage[T1]{fontenc}

\DeclareRobustCommand{\VAN}[3]{#2}
\let\VANthebibliography\thebibliography
\def\thebibliography{\DeclareRobustCommand{\VAN}[3]{##3}\VANthebibliography}

\usepackage[table]{xcolor}
\usepackage{lscape}
\usepackage{graphicx}	
\usepackage{amsmath}	
\usepackage{amssymb}	
\usepackage{natbib, bm, marvosym, enumitem, mathtools, epstopdf, color}  
\newcommand{\mstar}{{M}_{\star}}

\newcommand{\msun}{{\rm M}_{\odot}}

\newcommand{\kms}{~\rm{km}\cdot\rm{s^{-1}}}

\usepackage{newtxtext,newtxmath}

\title[Milky Way Mass from Gaia and Phat ELVIS]{Sizing from the Smallest Scales: The Mass of the Milky Way}

\author[M. K. Rodriguez Wimberly et al.]{M. K. Rodriguez Wimberly,$^{1,2}$\thanks{E-mail: wimberlm@uci.edu}\thanks{NSF MPS–Ascend Postdoctoral Fellow}
M. C. Cooper,$^{2}$
D. C. Baxter,$^{2}$
M. Boylan-Kolchin,$^{3}$
J. S. Bullock,$^{2}$
\newauthor 
S. P. Fillingham,$^{4}$ 
A. P. Ji,$^{5}$
L. V. Sales,$^{1}$
J. D. Simon,$^{6}$
\\
$^{1}$Department of Physics and Astronomy, University of California Riverside, 900 University Avenue, CA 92507, USA \\
$^{2}$Center for Cosmology, Department of Physics \& Astronomy, 4129 Reines Hall, University of California, Irvine, CA 92697, USA \\
$^{3}$Department of Astronomy, The University of Texas at Austin, 2515 Speedway, Stop C1400, Austin, TX 78712, USA \\
$^{4}$Department of Astronomy, University of Washington, Box 351580, Seattle, WA 98195, USA \\
$^{5}$Department of Astronomy \& Astrophysics, The University of Chicago, Chicago, IL 60637, USA \\
$^{6}$Observatories of the Carnegie Institution for Science, Pasadena, California 91101, USA
}

\date{Accepted 2022 May 3. Received 2022 May 3; in original form 2021 August 16}

\pubyear{2021}

\begin{document}
\label{firstpage}
\pagerange{\pageref{firstpage}--\pageref{lastpage}}
\maketitle

\begin{abstract}
As the Milky Way and its satellite system become more entrenched in near field cosmology efforts, the need for an accurate mass estimate of the Milky Way's dark matter halo is increasingly critical.
With the second and early third data releases of stellar proper motions from {\it Gaia}, several groups calculated full $6$D phase-space information for the population of Milky Way satellite galaxies. 
Utilizing these data in comparison to subhalo properties drawn from the Phat ELVIS simulations, we constrain the Milky Way dark matter halo mass to be $\sim 1-1.2\times10^{12}~\msun$. 
We find that the kinematics of subhalos drawn from more- or less-massive hosts (i.e.~$>1.2\times10^{12}~\msun$ or $<10^{12}~\msun$) are inconsistent, at the $3\sigma$ confidence level, with the observed velocities of the Milky Way satellites. 
The preferred host halo mass for the Milky Way is largely insensitive to the exclusion of systems associated with the Large Magellanic Cloud, changes in galaxy formation thresholds, and variations in observational completeness.  
As more Milky Way satellites are discovered, their velocities (radial, tangential, and total) plus Galactocentric distances will provide further insight into the mass of the Milky Way dark matter halo.
\end{abstract}



\section{Introduction}\label{sec:intro}

Several of the most pressing cosmological problems challenging the $\Lambda$CDM paradigm, namely the Too Big to Fail \citep[TBTF,][]{bk11,bk12} and the Missing Satellites problems \citep{moore99,klypin99}, depend heavily on the Milky Way's dark matter halo mass.
One way to resolve the TBTF problem within $\Lambda$CDM is through the assumption of a less massive Milky Way, for which fewer massive satellites with high central densities are expected \citep{Wang12, cautun14}.
Similarly, in conjunction with suppression of galaxy formation on the very smallest scales \citep[e.g.][]{efstathiou92,thoul96}, the Missing Satellites problem can also be largely eliminated by lowering the assumed Milky Way dark matter halo mass (and thus the predicted number of satellite systems). 
As such, the Milky Way's dark matter halo mass is a critical parameter in testing $\Lambda$CDM and models of galaxy formation on small scales \citep[see discussion in][]{Bullock2017}. 

Alternative resolutions to both the TBTF and Missing Satellites problems lie in the possibility that the Milky Way may be an outlier relative to the cosmic norm.
For example, only $\sim10\%$ of Milky Way-like systems are estimated to have satellites as massive as the Large and Small Magellanic Clouds \citep{bk10, busha11, tollerud11, santos21}.
The Milky Way's satellite population is also remarkable in another characteristic -- its Vast Polar Structure \citep[VPOS, e.g.][]{lb76, Kroupa05, Pawlowski12a, fritz18, Pawlowski20}.
While our ability to observe such structures in systems beyond our very local Universe is still relatively new \citep{Ibata13, Conn13, Collins15, Muller18}, our observed flattened polar distribution of satellites, the VPOS, seems to be uncommon \citep{Metz08, Pawlowski14a, Pawlowski14b, Ibata14, cautun15, Buck16, Ahmed17, Shao18, Shao19}.
Another unusual feature of our local system may be the high fraction of quenched (or passive) satellite galaxies.
Extragalactic surveys, such as SAGA, have found that the majority of satellites around Milky Way-like systems are actively star forming -- SAGA in fact finds $85\%$ of low-mass satellites ($\mstar \sim 10^{7 - 8.5}~\msun$) are star-forming across $36$ Milky Way-like systems \citep{mao20, geha17}.
The radial distributions of the Milky Way satellites versus observed and simulated Milky Way analogues is also a contentious point which may place the Milky Way out of the cosmic norm. 
Some recent work highlights discrepancies in the 3D radial distributions of Milky Way satellites and various cosmological simulations, in particular that Milky Way satellites are more radially concentrated than their simulated counterparts (\citealt{moore01, willman04, yniguez14, Carlsten20}, but see also \citealt{maccio10, Samuel20, font20, bose20}).
The dark matter halo mass of the Milky Way has strong implications for its ability to quench satellite galaxies and for the radial distribution of its satellite population.
More broadly, our reliance upon the Milky Way as a Cosmic Rosetta Stone \citep{bk16} requires a strong constraint on its dark matter halo mass.

{\it Gaia} has opened a new opportunity to study the distribution and dynamics of the Milky Way satellite population and to constrain the Milky Way's dark matter halo mass. 
Prior to the second data release (DR2) of proper motions from {\it Gaia}, the Milky Way's dark matter halo mass limits were $0.8 - 4.5 \times 10^{12}~\msun$ (e.g.~ \citealt{bk13, phelps13, kafle14}, and in particular see Figure $1$ in \citealt{Wang20} for a recent literature summary).
Since {\it Gaia} DR2, this mass has been inferred in various ways $-$ from calculating the escape speed from counter-rotating stars in the Galaxy’s outer halo \citep{monari18}; using a scale-free mass estimator involving the density, potential, and anisotropy ($\beta$) of the satellites $-$ galaxies or globular clusters $-$ surrounding the Milky Way \citep{watkins19, fritz20}; comparing phase-space distributions in simulations and semi-analytic models to observed distributions to then infer the mass \citep{patel18, eadie19, li20, callingham19}; fitting physically motivated models to the {\it Gaia} DR2 stellar rotation curve \citep{cautun20}; and calculating the mass within $100$ kpc via a distribution function method then extrapolating total mass \citep{deason20}.
The results of these recent studies range from $M_{200} = 0.7^{+0.11}_{-0.08}-1.55^{+0.64}_{-0.51} \times 10^{12}~\msun$ \citep[respectively]{eadie19, monari18}.
It is important to note that even within confidence intervals, many of these results do not agree with one another.

As an alternate approach to these direct dynamical methods, in this work we constrain the Milky Way's dark matter halo mass through comparison of subhalo kinematics in a suite of high-resolution $N$-body simulations to corresponding observational measures of the Milky Way satellite population using a statistical test to measure the {\it "anti–goodness of fit"} between these data sets.
Herein, we utilize orbital parameters for the Milky Way satellites, derived primarily from proper motion measurements contained in the early third {\it Gaia} Data Release \citep[EDR3,][]{gaiaedr3, gaiaproj}, subhalo kinematics from the Phat ELVIS suite of Milky Way–like simulations \citep{kelley19} and the Mann–Whitney {\it U} test \citep{mann47}.
In \S\ref{sec:data}, we discuss our observed tracers of the Milky Way host potential along with the comparison suite of cosmological simulations.  
\S\ref{sec:analysis} details our primary analysis techniques, while our results are presented in \S\ref{sec:results}.
In \S\ref{sec:discussion}, we examine various sources of potential systematic errors and points of further discussion, including the impact of satellites associated with the Large Magellanic Cloud, our adopted lower limit for peak subhalo velocity, orbital characteristics, limitations of our approach, comparisons to previous studies and observational completeness. 
Additionally, in this section we make some predictions for how future observations might impact our results.  
Finally, we summarize in \S\ref{sec:conclusion}.

\begin{figure} 
\centering
\includegraphics[width=0.45\textwidth]{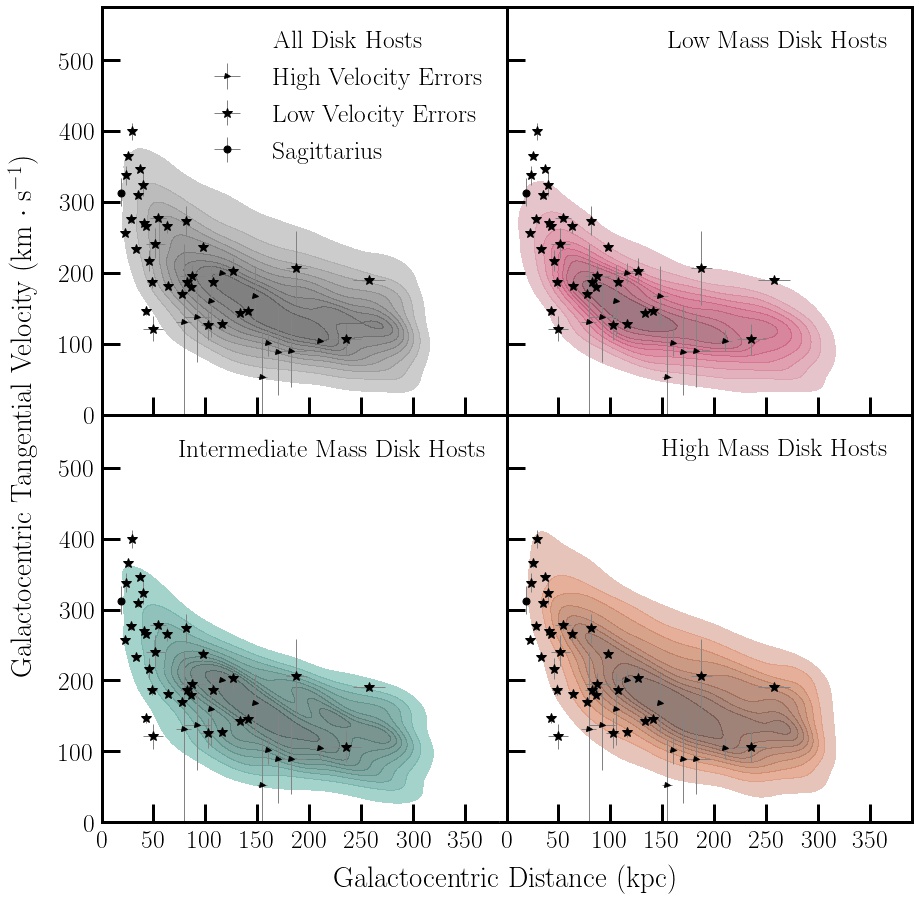}
\caption{Comparison of phase-space distributions between Phat ELVIS subhalos (density contours) and satellites in \citet{mcconnedr3} (MCV20a, black markers). In each plot, the stars denote satellites with low tangential velocity errors ($V_{\rm tan,err} \leq 0.30~V_{\rm tan}$). The triangles denote all other satellites excluding Sagittarius, which is represented as the circle. Sagittarius is poorly reproduced by the simulations and thus omitted from this analysis. The total velocity information for Sagittarius comes from \citet{fritz18}, which uses {\it Gaia} Data Release 2 proper motions. The total velocity errors are taken from MCV20a while the distance errors are taken from the literature (mainly \citealt{Simon2019} and references therein). The top left panel shows the phase-space density contours for subhalos across all $12$ Phat ELVIS hosts. Meanwhile, in the three remaining panels, we display the corresponding contours with subhalos divided according to host mass --- $0.7$--$1 \times 10^{12}~\msun$ (low mass, burgundy shading), $1$--$1.2 \times 10^{12}~\msun$ (intermediate mass, aqua shading), and $1.4$--$2 \times 10^{12}~\msun$ (high mass, sienna shading). Each mass bin includes $4$ hosts, with the adopted color scheme for the host mass sets carried throughout this paper.}
\label{fig:phasespace} 
\end{figure}

\section{Data}
\label{sec:data}

\subsection{\emph{Gaia}}
\label{subsec:gaia}

{\it Gaia} has spurred a dramatic improvement in our understanding of the orbital parameters for nearby stars, including those within the satellites of the Milky Way (MW) \citep{gaiaproj}.  
In the second and early third data releases \citep[DR2 and EDR3,][]{gaiadr2rel, lindegren18, gaiaedr3}, {\it Gaia} provides precise parallaxes and/or proper motions for over one \emph{billion} sources, in an absolute reference frame defined entirely by {\it Gaia} observations. 
From this vast data set, several groups calculated full phase-space information, including tangential velocities, for a majority of the MW satellites (e.g.~using DR2: \citealt{Helmi18, Simon2018, fritz18, massari18, Pace18, kallivayalil18} plus \citealt{mcconnedr3, li21, battaglia22}, which utilize EDR3).
Herein, we utilize the Galactocentric tangential velocities from \citet[][hereafter referred to as MCV20a]{mcconnedr3}.
Heliocentric radial velocities are taken from \citet{mcconndr2} and converted to the Galactocentric reference frame using \texttt{astropy} \citep{astropy13, astropy18}.
Heliocentric distances and associated errors are taken from \citet{Simon2019}\footnotemark, \citet{karachentsev06}, \citet{Weisz2016}, and \citet{Torrealba2019}.
These distances are also converted to the Galactocentric reference frame using \texttt{astropy}.
These quantities, converted to the Galactocentric reference frame, along with other properties of the MW satellies used in this work can be found in Table~\ref{table:1a}.

\footnotetext{Data presented by \citet{Simon2019} are compiled from \citet{Torrealba2016, DallOra2006, Walsh2008, Kuehn2007, Greco2007, Torrealba2018, Musella2009, Torrealba2016a, Kinemuchi2008, Longeard2018, crnojevic16, Rizzi2007, koposov15, Musella2012, des1, Koposov2018, Vivas2016, Bellazzini2004, Bellazzini2005, Moretti2009, Medina2018, sand12, Mutlu-Pakdil2018, Pietrzynski2008, belokurov07, Boettcher2013, Lee2003, hamanowicz16, Carlin2017, des2, Garofalo2013, Dallora2012, willman05a}.}

We limit our sample of Milky Way satellites to those systems within a Galactocentric distance of $300~{\rm kpc}$ and exclude unconfirmed systems that are likely not galaxies (e.g.~Indus~I and DESJ0225+0304). In addition, we exclude the Sagittarius (Sgr) dwarf from our sample. Sgr is currently being disrupted via tidal interactions with the Milky Way \citep[e.g.][]{ibata94, law09, koposov12}, such that it is poorly reproduced in our comparison simulation data set (see \S\ref{subsec:phELVIS}). %
Our primary sample includes $44$ satellite galaxies. 
Figure~\ref{fig:phasespace} shows the distribution of these systems as a function of Galactocentric distance and total velocity. 
Finally, we identify a subsample of $34$ systems with higher-precision tangential velocities, such that  $|V_{\rm{tan, err}} / V_{\rm{tan}}| \leq 0.30$. 
While this subset of systems is biased towards smaller Galactocentric distance, it does span a broad range of velocities (see Fig.~\ref{fig:phasespace}).
The particular selection limit used to define this subsample was adopted to exclude those systems with exceptionally uncertain tangential velocities while maintaining a statistically significant sample size.  

\subsection{Phat ELVIS}
\label{subsec:phELVIS}

As a comparison data set, we utilize the Phat ELVIS (phELVIS) suite of $12$ high-resolution, dissipationless simulations of MW-like halos \citep{kelley19}.
Building upon the ELVIS (Exploring the Local Volume In Simulations) suite of Local Group and MW-like simulations \citep{gk14}, phELVIS incorporates the effects of tidal disruption due to an artificial disk potential \citep[e.g.][]{gk17, sawala17}.
This new suite includes a total of $24$ MW-like simulations, encompassing $12$ high-resolution cosmological dark matter-only (DMO) simulations of isolated MW-like halos and $12$ re-runs of those DMO simulations with an embedded galaxy potential matching the observed MW disk and bulge (from here on referred to as the Disk runs). 
The $12$ Disk runs begin as identical duplicates to the $12$ DMO suites. 
At $z = 3$, a galaxy potential, including a stellar disk, gaseous disk, and Hernquist bulge component, is inserted into each of the Disk hosts.\footnote{At $z=0$ the inserted masses are $M_{\rm{Stellar~Disk}} = 4.1 \times 10^{10}~\msun$, $M_{\rm{Gas~Disk}} = 1.9 \times 10^{10}~\msun$, and $M_{\rm{Buldge}} = 0.9 \times 10^{10}~\msun$, see Table 1 of \cite{kelley19} for more details.}
While the potentials temporally evolve, each Disk host ends up at $z=0$ as an observationally-constrained MW \citep{kelley19}.

Each simulation occurs within a global cosmological box of length $74.06$~Mpc ($50~h^{-1}$~Mpc) with a dark matter particle mass of $3 \times 10^{4}~\msun$ and a Plummer-equivalent force softening length of $\epsilon = 37$~parsecs. 
These parameters allow for the subhalo catalogs to be complete down to a maximum circular velocity of $V_{\rm max} > 4.5~{\rm km}~{\rm s}^{-1}$, i.e.~a total bound mass of $\gtrsim 5 \times 10^{6}~\msun$. Refer to Figure 2 in \citealt{kelley19} for a visualization of completeness limits \textemdash~ in $V_{\rm peak}$ cumulative distributions, roll\textendash off of the functions begin at low $V_{\rm peak}$ (at approximately $V_{\rm peak} \approx 6~{\rm km}~{\rm s}^{-1}$ within $50$ kpc), yet throughout the radii explored, there is no roll\textendash off in $V_{\rm max}$ going as far down as $V_{\rm max} = 4.5~{\rm km}~{\rm s}^{-1}$. This serves as a measure of incompleteness.
The phELVIS halo catalogs are constructed of $152$ snapshots, evenly spaced in scale factor with a time resolution of roughly $100$~Myr. 
We spline interpolate the subhalo positions and velocities to achieve a time resolution of $\sim10$~Myr and use the interpolated data to calculate only pericentric passage and infall time.
Phat ELVIS adopts the cosmology of \citet{planck16} with the following parameters: $\Omega_{\rm m} = 0.3121$, $\Omega_{\Lambda} = 0.6879$, and $h = 0.6751$.

In order to more directly compare to the MW satellite population, we select subhalos from the phELVIS suite with $V_{\rm peak} > 6~{\rm km}~{\rm s}^{-1}$. 
While suppression of galaxy formation due to reionization is often predicted to occur below a mass limit of $V_{\rm peak} \sim 20-25~{\rm km}~{\rm s}^{-1}$ \citep[e.g.][]{gnedin00, hoeft06, ocvirk16}, the observed abundance of ultra-faint satellites of the MW are better matched via a lower mass limit \citep{graus18}.
Adopting a more inclusive mass selection yields a considerably larger subhalo population for comparison, better sampling the host potential and allowing control of systematics associated with observational completeness. 
Radial profiles and infall times are both potentially biased by $V_{\rm peak}$ limits. Recent work has shown that subhalo radial profiles are largely independent of $V_{\rm peak}$ limits \citep{Newton18} and investigating the infall–$V_{\rm peak}$ relation in phELVIS illustrated that there is a tight correlation amongst distributions of subhalo infall times throughout our $V_{\rm peak}$ range, where the median difference between distributions is $0.6$ Gyr.

While the phELVIS subhalos catalogs at $z = 0$ provide thousands of subhalos for comparison, they are limited to a single snapshot of each subhalo orbit. 
To better sample the host potential, we expand our subhalo population to include subhalos at two earlier snapshots. 
These two other snapshots were selected based on the average growth histories of the phELVIS hosts to minimize variation in the host mass. 
A majority of the hosts have minor ($\lesssim 2\%$) to no growth after $z=0.05$, which corresponds to the original $8^{\rm th}$ timestep prior to $z = 0$. The two snapshots chosen to examine here are evenly spread -- specifically the  $8^{\rm th}$ and $4^{\rm th}$ (corresponding to $z=0.05, 0.02$, respectively).
The vast majority ($\sim 90\%$) of the subhalo population is present at all 3 timesteps, with a small number of subhalos missing (or added) at earlier timesteps due to recent accretion, backsplashing, and/or tidal destruction.

Throughout this work, we focus on the $12$ MW-like hosts in the Disk runs.
The hosts with embedded disk potentials are chosen for their ability to better represent the observations relative to the dark matter–only (DMO) hosts. In the DMO runs, the greatest subhalo Galactocentric total velocities, which are all found at small Galactocentric distances, are systematically lower than those found in the Disk runs. This trend is seen in all three of the host halo mass ranges displayed in Fig.~\ref{fig:phasespace} and further strengthens the argument initially made in \cite{kelley19} — central–galaxy dynamics must be included to match observations of the satellite population.
The Disk host halos range in virial mass\footnotemark~from $0.71-1.95 \times 10^{12}~\msun$. \footnotetext{In the phELVIS simulations, virial mass, $M_{\rm vir}$, follows the \cite{bn98} definition.}
We split the hosts evenly into $3$ groups based on mass -- least massive, intermediate mass, and most massive. 
Specifically, the mass ranges of the $3$ bins are: $0.71-0.96 \times 10^{12}~\msun$ (low mass), $1.04-1.20 \times 10^{12}~\msun$ (intermediate mass), and $1.40 - 1.95 \times 10^{12}~\msun$ (high mass).
Excluding subhalos with $V_{\rm{peak}} < 6~{\rm km} {\rm s}^{-1}$, there are an average of $1200$ subhalos ($< R_{\rm vir}$) associated with each of the $4$ low-mass hosts, in comparison to an average of $1400$ ($2100$) subhalos for each of the $4$ intermediate-mass (high-mass) hosts.
Within each of the $3$ host mass bins, the halo-to-halo scatter in subhalo count is not great. The normalized, cumulative distribution of each bin is approximately the average of the $4$ individual host distributions that comprise that particular bin.
As shown in Figure~\ref{fig:phasespace}, due to the tidal disruption of subhalos in the Disk runs, phELVIS includes exceedingly few analogs to the Sagittarius dwarf. As discussed in \S\ref{subsec:gaia}, for this reason Sgr is excluded from the sample of MW satellites studied. 

\begin{figure} 
\centering
\includegraphics[width=0.5\textwidth]{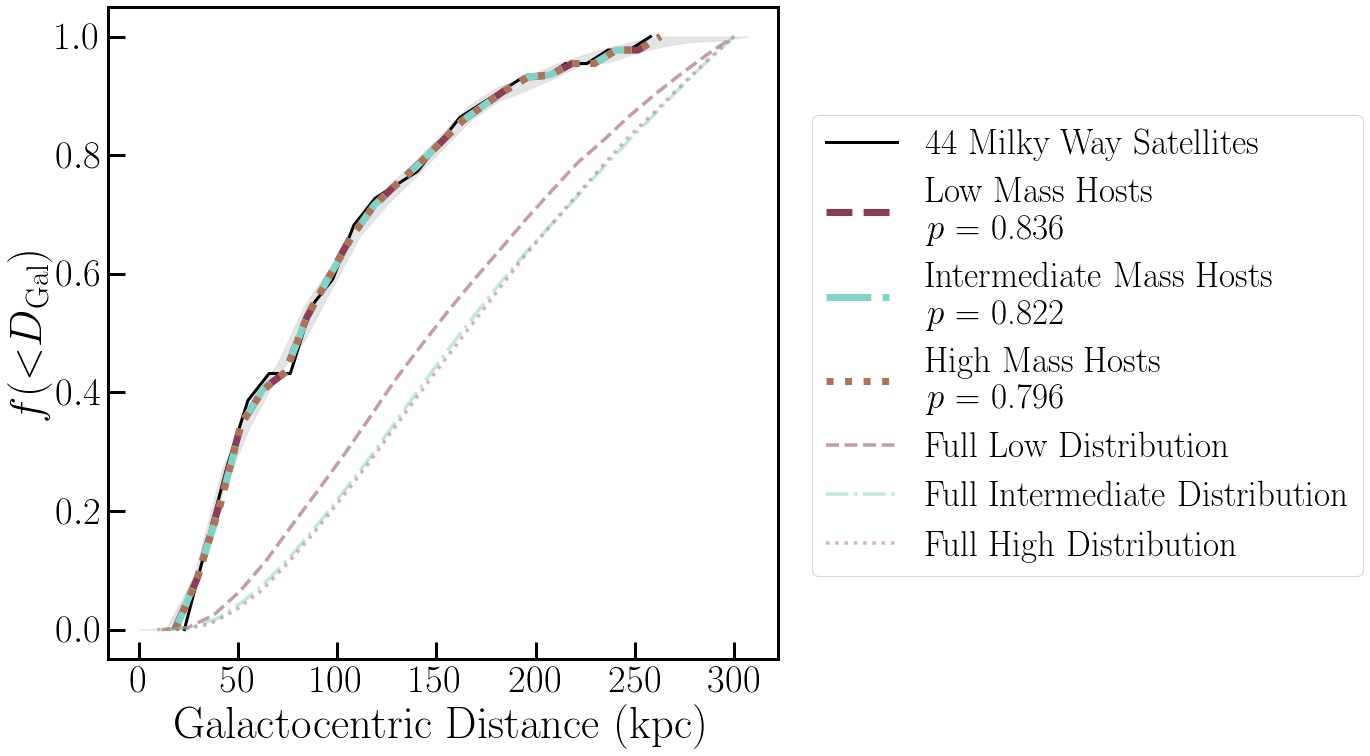}
\caption{The cumulative distributions of MW-centric and host halo-centric physical distances are engineered to be nearly identical via our distance-matching scheme. The solid black line is the cumulative distribution of Galactocentric distances for our primary sample of $44$ satellite galaxies. The dashed burgundy line, dash-dotted aqua line, and the dotted sienna line are the corresponding cumulative distributions for subhalos drawn from the three host halo sets, where Galactocentric distance is measured with respect to the corresponding host halo. The thinner dashed, dash–dotted and dotted lines are the cumulative distributions of all the subhalos in the respective host mass bins. The (thin) grey shaded region is the range of reported errors in the literature (see \citealt{Simon2019} and references therein).
The Mann-Whitney U Test $p$-value statistic is calculated using these errors. The legend reports the harmonic mean of MWU $p$-values from 500 randomly-selected, distance-matched subhalo distributions. This tight comparison (as we cannot reject the null hypothesis that the distributions are drawn from the same parent population due to the two-sided $p$-values all being well above the statistical significance level of $0.05$), reduces the potential biases associated with the incompleteness of the MW satellite population.}
\label{fig:dist} 
\end{figure}

\section{Analysis}\label{sec:analysis}

To study how the dynamics of the MW satellites depend on host halo mass, we select subhalos from the phELVIS simulations from each of the the host mass divisions discussed in \S\ref{subsec:phELVIS}. We then compare the subhalo samples to the observational data set via Galactocentric velocities -- namely, radial, tangential, and total.  
To mitigate selection effects driven by incompleteness in the sample of MW satellites, we match our sample of MW satellites to phELVIS halos via Galactocentric distance.
For each of the three sets of host halos, we randomly select (with replacement) $10$ subhalos for each satellite, selecting the subhalos from distance bins of width $10~{\rm kpc}$ centered on the Galactocentric distance of the satellite, where there is an average of $600$ subhhalos in each satellite's distance bin.
The size of the distance bins encompasses the observational radial distance errors for $\sim70\%$ of the MW satellites.
These distance-matched subhalos are randomly selected from the parent catalog that combines the subhalo populations from all three timesteps.
For our primary sample of $44$ satellites, this produces three comparison samples of $440$ halos each, associated with the low-mass, intermediate-mass, and high-mass hosts.
As shown in Figure~\ref{fig:dist}, this distance-based matching enforces a very close correlation between the Galactocentric distances of the MW satellite sample and our comparison subhalo samples.   
One caveat to this method of matching -- it does not guarantee each distance-matched halo is located within $R_{\rm vir}$ of the host.
For example, Leo~I's Galactocentric distance is $258~{\rm kpc}$ while roughly only $25\%$ of the explored
phELVIS host halos have virial radii greater than this distance. 
Thus, some of the subhalos, drawn from the lower-mass hosts and distanced-matched to Leo~I, may reside beyond $R_{\rm vir}$.
We choose not to match subhalos to observed systems on normalized distance (i.e.~Galactocentric distance which has been normalized to the MW/host's virial radius) to avoid introducing biases associated with the boundedness of a system at or near the virial radius or possible tidal disruption for systems near the host halo's center.

To quantify the observation--to--simulation comparisons (i.e.~to measure if the distanced-matched halos do not represent the MW satellites, an {\it "anti–goodness of fit"}), we employ the Mann-Whitney $U$ (MWU) test \citep{mann47, scipy}.
This is a non-parametric statistical ranked summation test that examines two independent samples. 
This test does not require any knowledge of the underlying distribution in either of the independent samples. 
To avoid possible underlying biases in the ranked summation \citep{fong19}, we increase our observational sample size by randomly sampling the observational quantities' errors to match the size of the distance matched halo population (i.e.~440 observational values are compared to 440 simulated values).
For satellite characteristics that have asymmetric error distributions (e.g.~for $V_{\rm tan}$ and $V_{\rm tot}$), the errors are drawn equally from the positive and negative sides. 
No galaxy in our observational set has plus-minus errors that are extremely different from one another.
Fortunately, the MWU is attuned to only median changes, compared to say the Kolmogorov-Smirnov (KS) test which is sensitive to the shape of the underlying distributions as well as the medians. 
The null hypothesis for the MWU is that the two independent samples are, in fact, drawn from the same parent distribution. 
We report the MWU test results as the associated two-sided $p$-values. 
Our statistical significance level to reject the null hypothesis is set at $p \leq 0.05$.

In an additional step towards bias avoidance, we conduct Monte Carlo sampling, by randomly selecting input values for the MWU test from the observational errors, $500$ times per parameter.
This is to reduce the possibility of sampling a randomly skewed distribution, and incorporate the observational errors into our modeling.
In the two–sided MWU test, we use $500$ sets of $10N$ distance-matched subhalos, which are randomly selected from $10$ kpc distance bins (which encompass the radial distance errors for most satellites), and $10N$ measures of corresponding satellite properties, as drawn from the observed error distributions. A $p$–value is calculated for each MWU test.
We then take the harmonic mean of these $500$ $p$-values and use this as our statistical result.

To create a baseline to our distance-matched analysis, we compare the unmatched subhalo distributions in the same way as comparing the matched distributions. Here, we compare $10 N$ subhalo properties randomly chosen from any subhalo in the host halo set (independent of distance) to a set of $10 N$ values randomly sampled from the error distributions of the observational quantities. We do this twice -- once where $N = 44$ for the full set of satellites and $N = 34$ for the set of systems with proportionally low tangential velocity errors. Finally, we create the random samples $500$ times, comparing the unmatched set to the observational set each time and then take the harmonic mean of the resulting $p$-values.
As detailed in Table~\ref{table:2}, the comparison of Galactocentric distances for both observational sets to each of the $3$ unmatched host halo sets are rejected at greater than $5\sigma$, where $5\sigma$ maps to $p = 0.00001$ as determined by our choice of significance level ($p = 0.05$) and the fact that we calculate a two--sided $p$--value.
The tangential and total velocity comparisons are also rejected at greater than $5\sigma$ for all three sets.
The radial velocity $p$-values are a bit different in these comparisons. For the low- and intermediate-mass host halos sets, the $p$-values are rejected at greater than or near $3\sigma$, with $3\sigma$ mapping to $p = 0.0027$. 
The comparison for the high-mass host halo set in radial velocity cannot be rejected ($p > 0.05$) -- this is the only non-rejectable null hypothesis between the unmatched subhalo distributions and the observations.

\begin{figure}
\centering 
\includegraphics[width=0.45\textwidth]{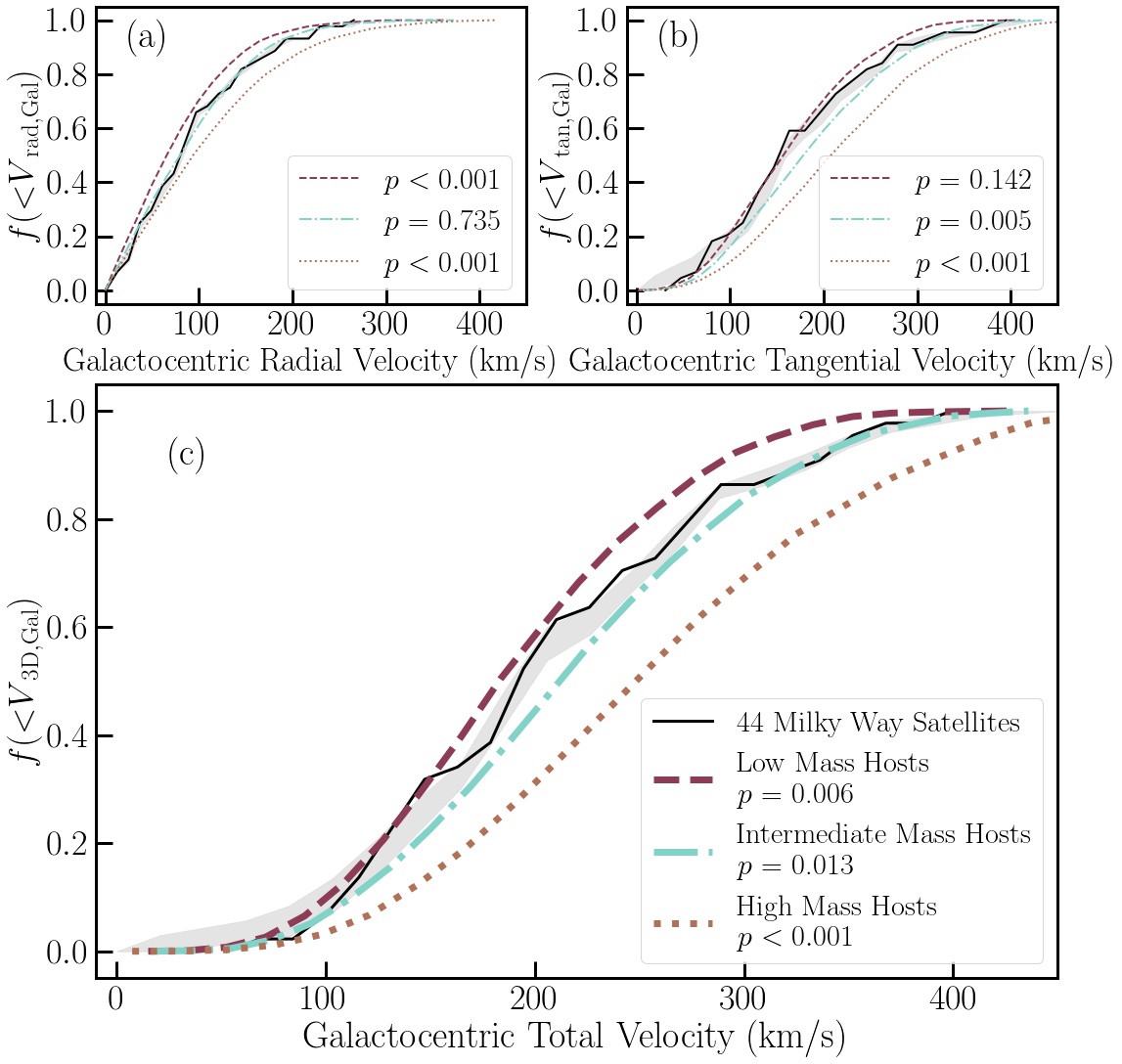}
\caption{Cumulative distributions of Galactocentric velocities $-$ namely in panel (a) $V_{\rm rad}$, in panel (b) $V_{\rm tan}$, and in panel (c) $V_{\rm tot}$ $-$ for the MW satellites in comparison to that of the simulated subhalos. The solid black line is the distribution for the $44$ satellite galaxies in the MCV20a sample. The dashed burgundy line, dash-dotted aqua line, and the dotted sienna line are the distributions for subhalos drawn from the three bins in host mass. The grey shaded regions are the cumulative distribution of the range of 500 randomly sampled values from each systems reported errors. In panel (a), the comparison in the well constrained parameter of Galactocentric radial velocity results in the preference towards only an intermediate mass MW dark matter halo. Panels (b) and (c) tell a different story. These subplots display how the inclusion of the further phase-space information from {\it Gaia} shifts the preference toward a less massive MW in Galactocentric tangential velocity, in that the low-mass hosts yield the only non-rejected $p$-value, while no host halo mass range is consistent with the observed Galactocentric total velocities for this sample.}
\label{fig:ns_allvel} 
\end{figure}

\section{Results}\label{sec:results}

We refine halo mass constraints for the MW by comparing the MW satellites’ Galactocentric velocities to distance-matched distributions of Phat ELVIS subhalos split into 3 groups based on host halo virial mass. 
The $3$ host halo mass bins range from $<10^{12}~\msun$ to $\sim2\times10^{12}~\msun$, with $\sim10^{12}~\msun$ being the intermediate bin. 
We focus on two sets of satellites drawn from the MCV20a sample --- all MW satellites and satellites with proportionally small tangential velocity errors (see~\S\ref{subsec:gaia}). These two subsets included $44$ and $34$ satellites, respectively.  

The subhalo distribution from each of the $3$ host halo sets is well matched in distance to each of the $2$ main satellite sets, by design (see Fig.~\ref{fig:dist} and \S\ref{sec:analysis}).
Given these subhalo samples that are well matched on Galactocentric distance to the observed MW satellite population, the velocity distributions of each satellite set is then compared to the corresponding measure for the distance-matched subhalos from the $4$ highest-mass host halos, the $4$ intermediate-mass host halos, and the $4$ lowest-mass host halos. 
This results in $3$ harmonic mean $p$-values for the each of the velocity components. 
When $p \leq 0.05$, the MWU test's null hypothesis can be rejected, which equates to the galaxy sample being poorly represented by the distance-matched subhalos in a specific host halo mass set (based on that particular measure of velocity). 

With tight distance-matched populations, we first examine the most well-constrained kinematic property -- Galactocentric radial velocity ($V_{\rm rad}$). 
As illustrated in panel (a) of Figure~\ref{fig:ns_allvel}, radial velocities for the main set of $44$ satellites are in good agreement with the corresponding velocities for subhalos in the intermediate-mass host halo set, while the subhalos drawn from the low-mass and high-mass hosts are inconsistent with the observations at $\gtrsim 5\sigma$ ($p < 0.00001$).

Since the strength of {\it Gaia}'s data is the ability to calculate full $6$-dimensional phase space, we take the analysis a step further by incorporating the not as richly studied Galactocentric total and tangential velocities.
For the main set of $44$ satellites, the preference for an intermediate-mass host halo is {\it not} evident when examining either of these two velocities -- i.e.~the associated $p$--values reject the null hypothesis that the two samples are drawn from the same parent distribution.
As shown in panels (b) and (c) of Figure~\ref{fig:ns_allvel}, the subhalos of the intermediate-mass hosts are inconsistent with the observational set.
In panel (b) the tangential velocity comparison prefers the low-mass hosts, i.e. this is the only non-rejected $p$-value, while in panel (c) the total velocity comparison rules out all host mass ranges -- i.e.~all samples yield $p < 0.05$.

A caveat to the {\it Gaia}-derived velocities is that a significant group of the observed systems have proportionally large errors associated with their proper motions which translates to proportionally large errors associated with the system's tangential and total velocities. 
These larger uncertainties allow for the possibility of extreme velocities that are not well represented in the phELVIS simulations. 
Inclusion of these systems in the analysis potentially creates a bias primarily towards lower-mass hosts. 
The systems in the MCV20a sample with proportionally high tangential velocity errors have the lowest tangential velocities of all $44$ satellites and are at distances further than roughly half the sample.
These kinematically cool systems become more rare with increasing host halo mass -- in higher-mass hosts, hotter systems are the norm.
For example, Leo~IV is one such system with proportionally large tangential velocity errors (i.e.~$|V_{\rm tan,err}/V_{\rm tan}| \ge 0.30$). Of the $1108$ halos within $\pm 5$~kpc of Leo IV's distance ($154.59 \pm 4.99$ kpc) in the $4$ highest-mass host halos, there are exactly $0$ halos with a tangential velocity in the bottom range of Leo IV's $1\sigma$ tangential velocity error ($V_{\rm tan} < 14~{\rm km}~{\rm s}^{-1}$). 

\begin{figure*} 
\centering 
\includegraphics[width=1.0\textwidth]{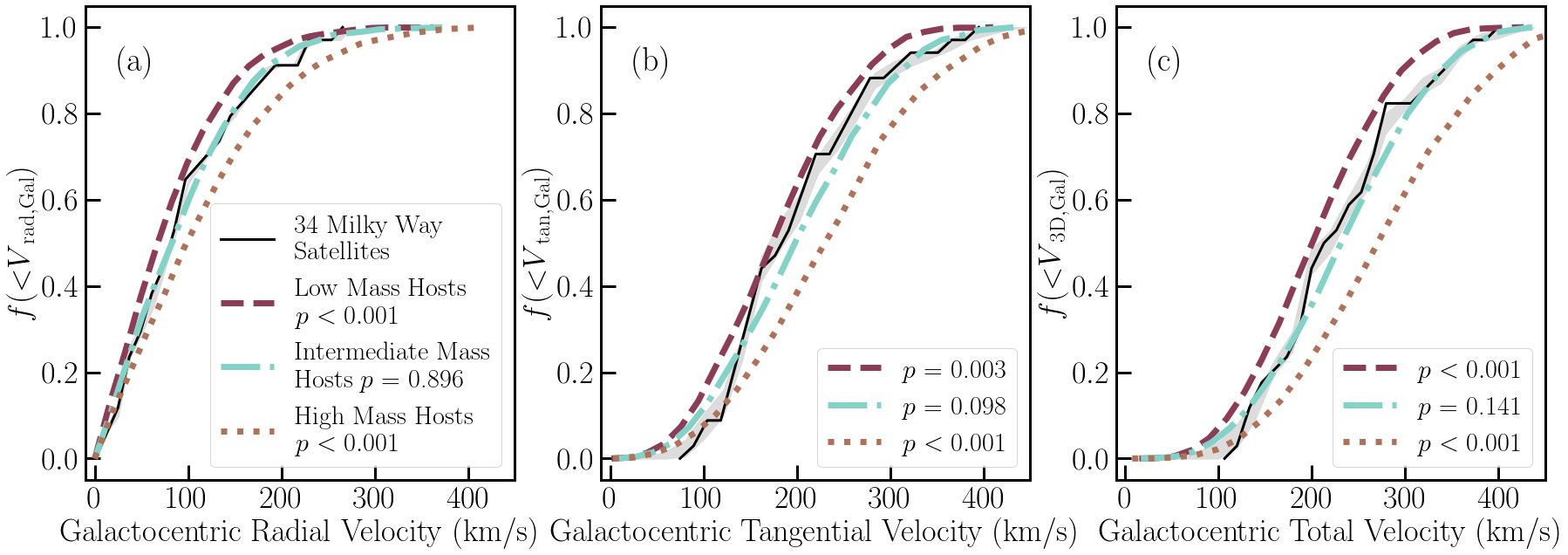}
\caption{Cumulative distributions of all velocity components for the pared-down set of $34$ satellites all with proportionally low tangential velocity errors. The color coding, line style and legend conventions are identical to Fig. \ref{fig:ns_allvel}. These three plots display how excluding satellites with their average plus-minus error $\ge30\%$ of their tangential velocity settles the host mass preference on an intermediate mass MW. Specifically, compared to Fig. \ref{fig:ns_allvel}, the resulting $p$-values for the intermediate mass host halos in the Galactocentric tangential and total velocities are now above the statistical significance level of $0.05$ while the $p$-values for the low- and high-mass host halos are rejected and therefore the intermediate mass hosts halos is the preferred host mass range.}
\label{fig:lvterr_vels} 
\end{figure*}

To address this potential bias, we rerun our distance-matching analysis using the subsample of $34$ satellites with low fractional uncertainty in $V_{\rm tan}$ -- specifically, we define proportionally low–error systems to have $|V_{\rm tan,err}/V_{\rm tan}| \le 0.30$. 
We then compare the resulting velocity distributions for this pared-down set.
As illustrated in Fig.~\ref{fig:lvterr_vels}, the radial velocity comparison, panel (a), is essentially unaffected by the removal of systems with proportionally high tangential velocity errors.
However, the comparison in tangential velocity space, panel (b), now prefers the intermediate-mass hosts.
The preferred halo mass in the total velocity comparison also changes from Fig.~\ref{fig:ns_allvel} to now prefer the intermediate-mass hosts. 
The low- and high-mass hosts in all $3$ velocity component comparisons are rejected ($p < 0.05$) at or near $3\sigma$.
All $p$-values discussed here can be found in Table \ref{table:2}.

To explore the limits of this preferred intermediate-mass range, $1.04-1.20 \times 10^{12}~\msun$, we rerun the analysis with thinner and wider intermediate-mass ranges. More specifically, we ran the analysis with a thinner intermediate mass range of $\sim1.04-1.10 \times 10^{12}~\msun$ ($3$ host halos) and a wider intermediate mass range of $\sim0.96-1.40 \times 10^{12}~\msun$ ($6$ intermediate-mass host halos instead of $4$). We compare these varying host halo mass ranges to the observational set of systems with proportionally low tangential velocity errors.
In the velocity comparisons for the thinner intermediate-mass host halo set, this new intermediate-mass host halo set of $3$ halos is preferred across all $3$ velocity components. The $3$ $p$-values for this host halo set are all greater than the significance limit of $0.05$, while the $p$-values for the low- and high-mass hosts in all $3$ velocity components are rejected at or near $3\sigma$.
In the velocity comparisons for the wider host mass set, there is not one host halo set preferred across all $3$ velocity components. Though the intermediate-mass sample is nearly preferred across the components -- the radial and total velocity $p$-values are greater than the significance limit while the tangential velocity $p$-value is just below this limit ($p = 0.045$).
\emph{Overall, the results based on comparing the set of MW satellites with proportionally low tangential velocity errors to the phELVIS simulations strongly indicate that the {\it Gaia}-based distances and velocities are consistent with a MW dark matter halo mass of $\sim 1 - 1.20 \times 10^{12}~\msun$.} 

\begin{figure}
\centering 
\includegraphics[width=0.45\textwidth]{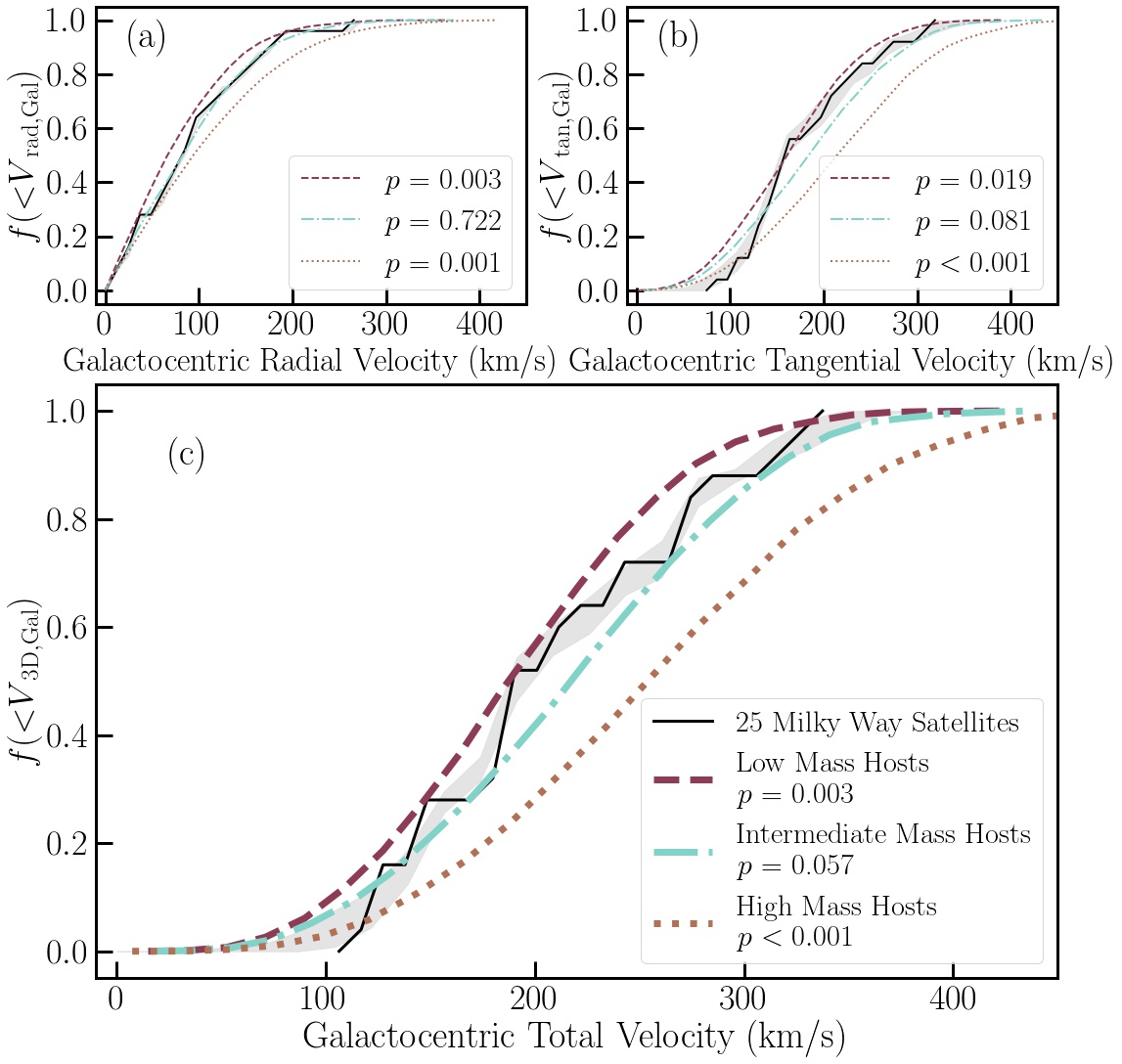}
\caption{Cumulative distributions of individual Galactocentric velocity components for the set of ($25$) satellite galaxies excluding all $9$ LMC-associated systems (and excluding satellites with proportionally high tangential velocity errors). The color coding, line style, and legend conventions are identical to Fig.~\ref{fig:ns_allvel}. 
The exclusion of the $9$ LMC-associated systems does not change the intermediate host mass preference. Across all $3$ velocity component comparisons, the $p$-values for the low- and high-mass host halos are rejected ($p < 0.05$) while the intermediate-mass host halo $p$-values are above the significance limit.}
\label{fig:lmc_vel} 
\end{figure}

\section{Discussion}\label{sec:discussion}

\subsection{Large Magellanic Cloud (LMC) Satellites}\label{ssec:lmcsat}

Of the $12$ MW-like systems in the phELVIS suite with an embedded disk potential, there is only $1$ host with a Large Magellanic Cloud-like subhalo (i.e.~with $M_{\rm vir} \ge 8 \times 10^{10}~\msun$). 
If this restriction is lowered to $M_{\rm vir} \ge 3 \times 10^{10}~\msun$, then there are $2$ Large Magellanic Cloud-like subhalos throughout the $12$ disk hosts. 
It can be argued that including MW satellites that were originally LMC satellites in our analysis may bias our results, as these systems may not be  well-represented in the phELVIS simulation suite. 
Here, we explore the impact of removing LMC satellites from our observational sample of galaxies with low tangential velocity errors.

While there is some contention over which galaxies are satellites of the LMC, 
the derived proper motions from {\it Gaia} DR2 have allowed for more direct investigation into potential associations with the LMC.
In addition to Horologium I, which has been found to be a likely LMC satellite in multiple studies \citep{sales17, erkal19, patel20, santos21}, Carina II, Carina III, and Hydrus I have also been classified as long-term satellites of the LMC via their {\it Gaia} DR2 proper motions \citep{kallivayalil18, patel20}, where long-term is defined by being bound to the LMC for at least $2$ consecutive orbits. 
Furthermore, \citet{patel20} found another $5$ galaxies to be recently-captured LMC satellites (Reticulum II and Phoenix II) or have had prior interactions with the LMC (Sculptor, Segue 1, and Tucana III).

To explore how our results may be biased by the dynamical influence of the LMC, we fully rerun our distance-matching analysis on the set of MW satellites with low tangential velocity error fractions excluding the 9 LMC-associated satellites, which includes long-term satellites plus recent satellites and LMC interactors.
When excluding these systems associated with the LMC, across all $3$ velocity components, the null hypothesis is rejected when comparing to subhalos drawn from the low- and high-mass hosts (i.e.~$p < 0.05$), with the distribution of observed velocities for the Milky Way satellites intermediate between these two subhalo samples (i.e.~again favoring an intermediate-mass Milky Way).
However, while the radial velocity distribution for subhalos drawn from the intermediate-mass sample is  visually consistent with that of the Milky Way satellite population when excluding the LMC-associated satellites, we find that the tangential and total velocities are less consistent (as compared to the distributions in Fig.~\ref{fig:lvterr_vels}).
As seen in Fig.~\ref{fig:lmc_vel} and Table~\ref{table:pvales}, the intermediate-mass host samples yield non-rejected $p$-values, $p \ge 0.05$, when comparing to the observed radial, tangential and total velocities. 
Overall, when satellites associated with the LMC are removed from our analysis, the distance-matched subhalos continue to show a preference for an intermediate-mass host halo ($\sim 1-1.2 \times 10^{12}~\msun$).

So far in this work, all distance-matched subhalos are selected according to a peak maximum circular velocity limit of $V_{\rm peak} > 6~{\rm km}~{\rm s}^{-1}$. 
As shown by \citet{graus18}, the abundance of Milky Way satellites can be reproduced with subahlos down to $V_{\rm peak} >10~{\rm km}~{\rm s}^{-1}$ when excluding those systems associated with the LMC. 
While this more restrictive subhalo selection roughly quarters the subhalo populations, only Draco~II does not have at least $10$ subhalos in its distance bin across all host mass sets. 
As shown in Table~\ref{table:2}, when limiting subhalos to $V_{\rm peak} > 10~{\rm km}~{\rm s}^{-1}$, there is no consistent host mass preference. The high-mass host samples are inconsistent with the observed Milky Way satellites across all $3$ velocity measures. Meanwhile, the subhalos drawn from the low- and intermediate-mass hosts are consistent with the observed radial velocities, but unable to reproduce the tangential and total velocity distributions of the Milky Way satellites. 

\begin{figure*} 
\centering 
\includegraphics[width=1.0\textwidth]{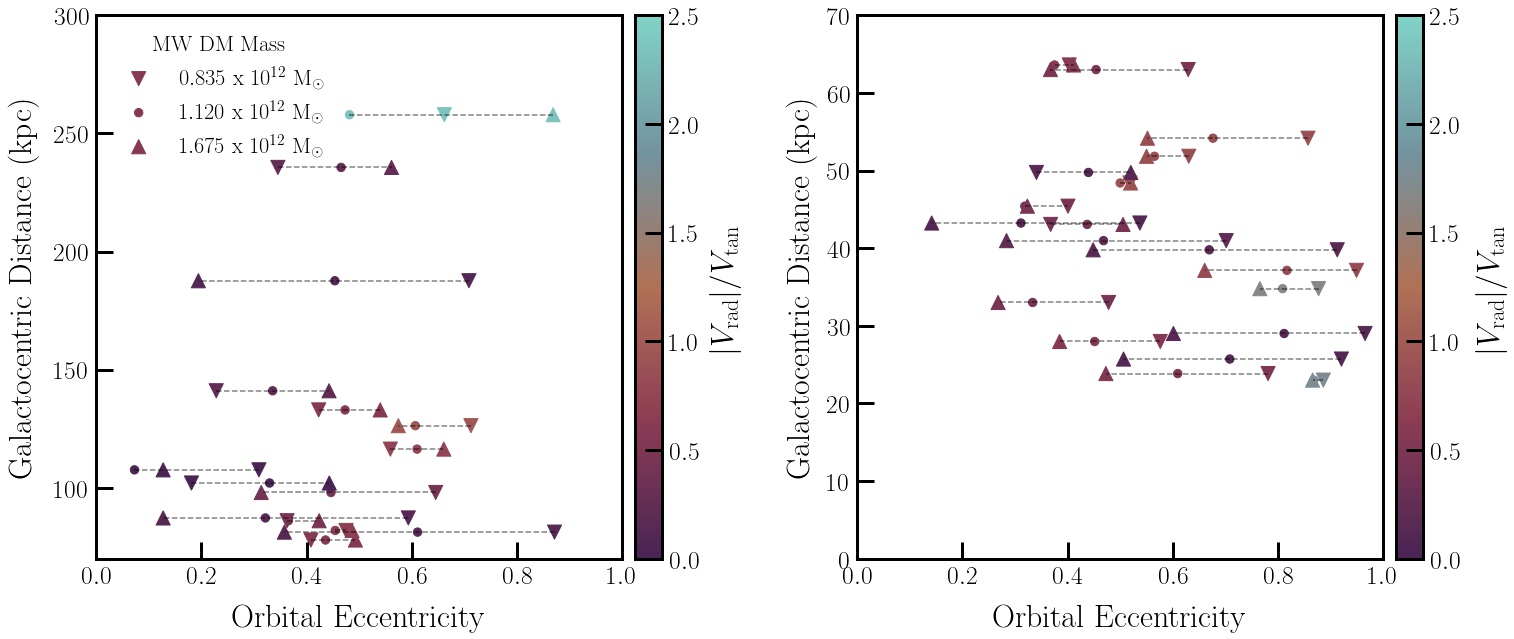}
\caption{Galactocentric distance against orbital eccentricity for the MW satellites with proportionally low tangential velocity errors color coded by the ratio of Galactocentric radial velocity versus tangential velocity. The satellites are separated into two groups: {\it (left panel)} systems outside $70$ kpc and {\it (right panel)} systems inside $70$ kpc. These figures display trends in eccentricity $-$ decreasing eccentricity with increasing MW dark matter halo mass for those systems inside $70$ kpc {\it (right panel)} and the lack of a strong trend in eccentricity for those systems outside $70$ kpc {\it (left panel)}. As seen by the color coding, $|V_{\rm rad}| / V_{\rm tan} > 1$ indicates more radial orbits while those $< 1$ point towards more circular orbits. Eccentricity may serve well as host mass diagnostics.}
\label{fig:lvterr_peri} 
\end{figure*}

\subsection{Pericentric Passage, Eccentricity \& Satellite Infall}\label{ssec:orbits}

To explore the orbits of the observed MW satellites in various dark matter halo potentials, we calculate pericenters and orbital eccentricities, defined as eccentricity $= (r_{\rm apocenter} - r_{\rm pericenter}) / (r_{\rm apocenter} + r_{\rm pericenter})$, for a low-, intermediate-, and high-mass MW potential using {\sc galpy} \citep{bovy15}. We employ {\sc galpy}'s standard \texttt{MWPotential2014} potential model, which contains a spherical buldge, a Miyamoto Nagai Potential disk and an NFW dark matter halow with concentration 15.3, along with the EDR3 proper motions and other satellite properties from the literature (i.e.~RA, Dec, heliocentric distance, and line-of-sight velocity). We modify the \texttt{MWPotential2014} by adopting $3$ host potentials based on the average halo masses from our $3$ host sets --- i.e.~average masses of $0.835$, $1.120$, $1.675 \times 10^{12}~\msun$ \textemdash ~$1.04$, $1.4$ and, $2.09\times$\texttt{MWPotential2014}, respectively. The resulting inferred orbital properties ($3$ pericenters and $3$ eccentricities) for the MW satellites can be found in Table \ref{table:1b}.

Comparing the resulting $3$ pericenteric passages derived using {\sc galpy}, there is a mild preference for smaller pericenters in increasing host potentials -- e.g.~for satellites $30-60$~kpc from the center of the Milky Way, the median pericenter in the largest host potential is $\sim 30$~kpc, while the median in the smallest potential is $\gtrsim 35$~kpc. This host mass$-$pericenter correlation is not as strong amongst the distance$-$matched phELVIS subhalos, where the median pericenters in this same distance bin ($30-60$~kpc) is roughly half the spread of that seen in the {\sc galpy} pericenters.\footnote{The subhalo pericenters are likely overestimated, as they were calculated after spline interpolating distances and velocities for each subhalo \citep{richings20}, which may affect the varying trends between data sets.}
While these predicted pericenters for all $44$ satellites decrease with increasing host potential, the correlation between orbital eccentricity and host halo mass is more complicated. 
As illustrated in Fig.~\ref{fig:lvterr_peri}, for satellites currently in the outer MW halo ($70 < D_{\rm MW}/{\rm kpc} < 300$), there is not a strong correlation between host potential and orbital eccentricity. In the inner MW halo ($D_{\rm MW} < 70$~kpc), however, there is a clear negative correlation, such that eccentricity decreases with increasing host potential for $\sim75\%$ of the satellites --- i.e.~orbits become more circular in greater potentials.
The preferential circularization of orbits with increasing host mass is likely the result of satellite disruption associated with tidal forces. At a given host-centric distance within the inner halo, surviving satellites of more massive hosts tend to populate circular orbits, as tidal destruction has preferentially destroyed systems on more plunging orbits. 

\begin{figure} 
\centering
\includegraphics[width=0.45\textwidth]{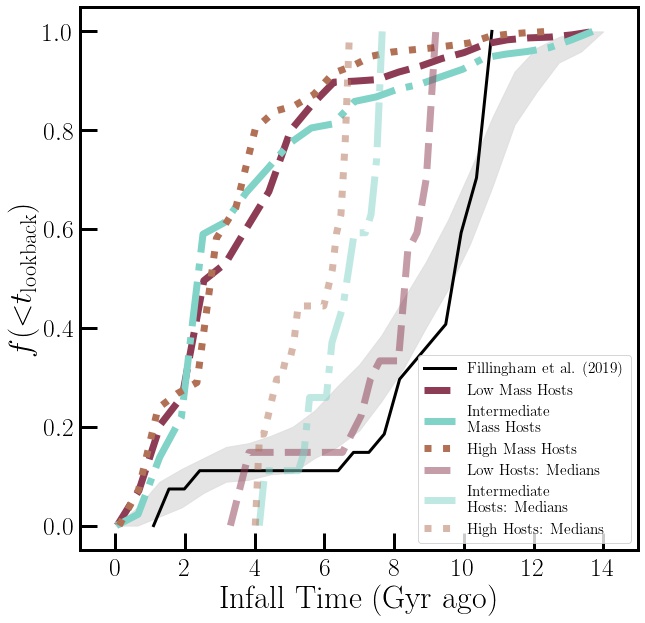}
\caption{Cumulative distribution of satellite/subhalo infall times. The solid black line shows the cumulative distribution of first infall times for $26$ MW satellites with low tangential velocity errors, as inferred by \citet{fham19}. The grey shaded region is the cumulative distribution of the range of 500 randomly sampled values from each of the $26$ systems' reported errors. The opaque dashed burgundy line, opaque dash-dotted aqua line, and the opaque dotted sienna line are the corresponding distributions for all the distance-matched subhalos belonging to our adopted host-mass bins. The half–transparent versions of these colorful lines are cumulative distributions of the median infall time for each set of subhalos distance–matched to each galaxy in the Fillingham et al. sample. These median distributions mimic methods employed by Fillingham et al. showing better agreement to their distribution. 
In contrast, the opaque lines account for the full range of possible infall times which allows for later infall times to become more common, skewing those distributions to earlier cosmic times.
As a whole, the distribution of infall times, as inferred from {\it Gaia} proper motions, does not match that found in the simulations, with the simulations favoring later (i.e.~more recent) infall times.}
\label{fig:lvterr_inf} 
\end{figure}

Beyond pericenter and eccentricity, another critical orbital parameter is the infall time onto the Milky Way (or host halo). 
Within the simulations, we are able to directly trace the infall of subhalos, such that infall time is defined as the lookback time when a subhalo \emph{first} crossed the host halo's virial radius. 
For observed satellites of the MW, on the other hand, constraining the infall time is more challenging (given that we lack a DeLorean and a flux capacitor).
Using {\it Gaia} proper motions from \citet{fritz18} to estimate the binding energy of each Milky Way satellite, \citet{fham19} estimate the infall time  according to a correlation between infall time and binding energy derived for subhalos in the phELVIS simulations \citep[see also][]{rocha12}. 
For each satellite's infall time, \citet{fham19} adopt the peak value in that satellite's kernel density estimation (KDE) from binding energy–matched subhalo infall times.
When computing the binding energy of the Milky Way satellites, they assume a host halo mass of $1.3\times10^{12}~\msun$, which is directly between that of our intermediate- and high-mass host halo samples. 
As illustrated in Fig.~\ref{fig:lvterr_inf}, however, the distribution of infall times within phELVIS is largely independent of host mass. 

For the $26$ MW satellites with proportionally low tangential velocity errors and infall times estimated by \citet{fham19}, we draw distance-matched subhalo samples as described in \S\ref{sec:analysis}. 
Figure~\ref{fig:lvterr_inf} shows cumulative distributions of infall times for all distance–matched subhalos in this subsample (opaque colorful lines). We note that the distribution of infall times obtained in this way is different than using only a single value, such as the the median of the distribution, for each set of subhalos distance–matched to each galaxy, which is shown by the half–transparent lines. Accounting for the full range of possible infall times (opaque lines) instead of just the median value per galaxy (half–transparent lines) allows for later infall times to become more common, independent of the host mass. Interestingly our distribution of median infall times (half–transparent lines) are in reasonable agreement with estimates from \citet{fham19}, which are based on energy–matching subhalos instead of distance–matching as in this work.

The inferred median infall times for the MW satellite population in phELVIS (half–transparent lines) are skewed to earlier cosmic times relative to the distribution of most likely (KDE peak) infall times calculated for the MW (black solid line), an effect even stronger when allowing for the whole distribution of infall times for the distance matched subhalos to be included (opaque lines). This suggests that it is possible that the MW may be an outlier with regard to its accretion history, such that a larger fraction of its satellites were accreted at early cosmic time \citep{elias18}, potentially via correlated accretion of substructures \citep{dsouza21}. While the likely recent infall of the LMC (and associated satellites) would counter this potential bias in accretion history to some degree \citep{besla07, kallivayalil13}, the \emph{possibly} anomalous satellite quenched fraction for the Milky Way might serve as further evidence of a bias towards early accretion -- and excess quenching -- relative to other nearby Milky Way-like systems \citep{wheeler14, fham15, geha17}. 

\subsection{Limitations and Comparison to Previous Studies}\label{ssec:prevlim}

One obvious limitation of our method is the underlying assumption that the formation history of the MW is represented in our sample of phELVIS simulations. Encouragingly, as discussed in Section \ref{ssec:lmcsat}, the low rate of large satellites with their own satellite systems ($>20\%$ phELVIS host halos host an MC system) does not affect our results. 
A limitation which does impact our results is the coarse mass determination due to small host halo sample size \textemdash~ i.e.~ phELVIS has only $12$ MW–like host halos. 
Due to time and computational costs of running new, larger simulations, doing so to obtain a more precise result is beyond the scope of this paper.

Historically there has been a wide range of estimates of the MW’s mass from $0.56 \pm 0.12 \times 10^{12} \msun$, via stellar stream modeling \citep[using $N$–body realizations of the Sagittarius stream]{gibbons14}, to $2.65^{+1.58}_{-1.36} \times 10^{12} \msun$, via timing mass argument \citep[using \textit{Hubble Space Telescope} (HST) derived proper motions for Leo I]{sohn13}. 
Within the uncertainties these results do not agree with one another or with our results. 
Recent studies using stellar streams, such as \cite{craig21}, who modeled the Magellanic Stream motivated by HST proper motions, found the mass to be $1.5 \pm 0.3 \times 10^{12} \msun$, which just overlaps our determination at $1\sigma$. 
The timing mass argument constrains the Local Group mass, and MW mass within it, through present–day kinematics, the impending major merger between the MW and M31 as well as mass restrictions to have overcome universal expansion (see \citet{blandhawthorn16, benisty20, benisty21} for thorough discussions on the method, its implications and limitations). 
Recent results using the timing argument, and data from the H3 Spectroscopic Survey, find a range from $0.9-1.5 \times 10^{12} \msun$ \citep{zaritsky20}, where our results fit snugly within.

There are many more approaches to calculating the virialized dark matter mass of the MW. 
Analyzing kinematics of various objects within the MW dark matter halo is one of the most common methods as there are many tracer object options (e.g. halo stars, globular clusters, hypervelocity stars, streams and satellite galaxies). 
The advent of {\it Gaia} has dramatically increased the precision of such efforts with its proper motion prowess. 
\citet{Wang20} found that of the studies basing their observational measurements off  {\it Gaia} DR2 data, the result of \citet{callingham19}, who compared satellite dynamics to model satellites in the EAGLE cosmological hydrodynamics simulations, is the median value at $1.17^{+0.21}_{-0.15} \times 10^{12} \msun$ \textemdash~ our results fully agree with theirs. 
The extremes of the {\it Gaia}\textendash based studies are $M_{200} = 0.7^{+0.11}_{-0.08} \times 10^{12}~\msun$ \citep[via cumulative mass profiles derived from globular cluster kinematics]{eadie19}, and $M_{200} = 1.31^{+0.45}_{-0.40} \times 10^{12}~\msun$ \citep[via satellite galaxy proper motions in a scale–free mass estimator]{fritz20}. Our results are in agreement with the more massive result but not with the least massive. 
As a caveat to kinematics–based mass estimations, \citet{erkal2020} recently illustrated that not including the LMC in such methods can result in overestimating the MW's mass by up to $50\%$ due to the LMC pushing the MW out of equilibrium. This issue is bypassed in methods directly comparing $N$–body simulations, where simulated systems are not in equilibrium, to the observed kinematics, as is done in this work.

As described in this section, the field is starting to converge on a well constrained value of the mass of the Milky Way where our results, $1-1.2 \times 10^{12} \msun$, fit within the range. 
Furthermore, the method presented here achieves a metric most other studies do not — we fully quantify which model masses deviate from the observed satellite population at $\gtrsim 3\sigma$ significance.
This ‘anti-goodness of fit’ measure is a great advantage of null hypothesis tests, such as the MWU.

\subsection{Observational Completeness}\label{ssec:obscomp}

With many new satellite galaxy discoveries within the past $2$ decades \citep{york00, willman05a, willman05b, zucker06a, zucker06b, belokurov10, des1, des2}, the debate related to the observational completeness of the MW satellite population has been revived \citep{Tollerud08, Walsh09, Wang13, Hargis14, gk17, Newton18, Jethwa18, Samuel20, Carlsten20}. 
By matching our subhalo subsamples to the observed MW satellite population based on host-centric distance, our analysis effectively minimizes any systematic bias associated with incompleteness. 
To more fully explore the potential impact of observational completeness on our results, however, we limit the Milky Way satellite population (and corresponding subhalo samples from phELVIS) to systems within $100$~kpc.
At these Galactocentric distances ($<100$~kpc), the MW satellite population is relatively complete, especially in the Southern Hemisphere thanks to surveys such as the Dark Energy Survey (DES) and other imaging campaigns using the Dark Energy Camera (DECam) on the Victor M. Blanco Telescope at the Cerro Tololo Inter-American Observatory \citep{dw20}. 

In our observational data set of systems with proportionally low tangential velocity errors, there are $25$ satellite galaxies within $100$~kpc. 
For this restricted -- yet largely complete -- sample, the observed velocity distributions are again inconsistent with the kinematics of subhalos drawn from the low- and high-mass host samples (see Table~\ref{table:2}). 
Meanwhile, while the distributions of observed radial velocities for the nearby Milky Way satellites and for the distance-matched phELVIS subhalos in the intermediate-mass hosts are consistent, the tangential (and total) velocity distribution for the nearby MW satellites is inconsistent ($p < 0.05$) with that of the intermediate-mass subhalo distribution. 
This slight disagreement between the velocity distributions is likely due to a preference for circular orbits at small host-centric distance in phELVIS, such that the distribution of $V_{\rm tan}$ is biased towards higher velocities relative to that of the observed MW satellites. This perhaps indicates that the tidal disruption of subhalos within phELVIS may be slightly over-estimated or otherwise incomplete in is characterization of orbits within the inner part of the host halo.  
\footnotetext{All $p$-values discussed in this section can be found in Table \ref{table:2}.}

\begin{figure*} 
\centering 
\includegraphics[width=1.0\textwidth]{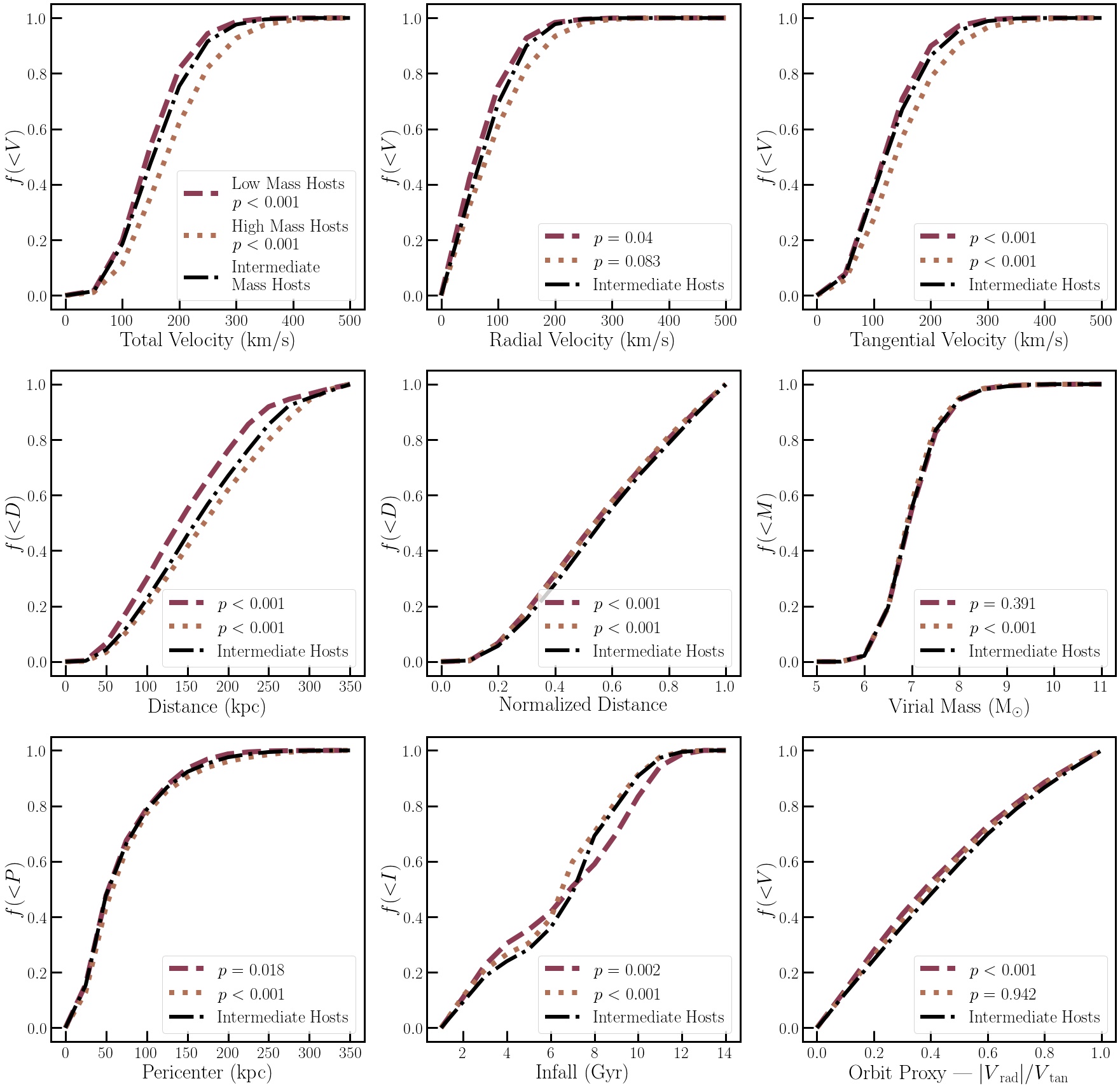}
\caption{Cumulative distributions of all halos within $350$ kpc of their respective host halo split into three host halo mass sets. The low mass (burgundy dashed lines) and high mass (sienna dotted lines) sets are compared to the intermediate mass (black dash-dotted lines) host halo sets across $9$ subhalo properties. These plots illustrate which physical parameters of newly discovered satellites will assist us in further refining the dark matter mass content of the MW $-$ namely two of the most straightforward to obtain properties -- distance and total velocity.}
\label{fig:prediction} 
\end{figure*}

\subsection{Observational Predictions}\label{ssec:obspred}
As more MW satellites are discovered through deep and wide imaging surveys, such as the Legacy Survey of Space and Time (LSST) at the Vera Rubin Observatory \citep{lsst} or the Nancy Grace Roman Space Telescope \citep{wfirst19}, or via future data releases from {\it Gaia}, the virial mass of the Milky Way might be further refined.
Fig.~\ref{fig:prediction} attempts to illustrate the potential future refinement based on different observable quantities. The cumulative kernel density estimates plotted trace the unmatched distribution of all halos within $350$~kpc of the respective host halo from hosts split into the three mass bins used throughout this work, 
with the median number of halos per host mass bin being 1650.
The low-mass (burgundy dashed lines) and high-mass (sienna dotted lines) sets are compared to the intermediate-mass (black dash-dotted lines) host halo sets across $9$ subhalo (or satellite) characteristics.
MWU $p$-values were calculated for comparisons between the subhalo distributions drawn from the low-mass and high-mass hosts relative to those in the intermediate-mass hosts.
Any characteristic with rejected $p$-values ($p < 0.05$), i.e. large differences between the three host mass binned subhalo populations, stand to be good metrics to test the preferred host mass range as new data becomes available.

Galactocentric total velocity, tangential velocity, physical Galactocentric distance, and infall have the most discernible differences between the host mass sets within phELVIS.
Since infall must be inferred from simulations \citep{fham19} or modeling the orbital history of the satellite \citep[e.g.][]{patel20} and thus has greater measurement uncertainty, it is likely to be of less help in discriminating between different host mass regimes. 
Distance and total velocity have the largest differences between subhalo distributions. Subhalos in the more massive hosts are kinematically hotter and at further distances from the center of their host halo than those in the less massive hosts.
As new satellites are discovered, the characterization of the radial selection function of observed MW satellites will improve, enabling distance to be used as a mass estimator. It will be particularly interesting to explore their total velocity\textendash based phase space to further refine the halo mass of the Milky Way. 

\section{Summary}\label{sec:conclusion}

Using the Phat ELVIS suite of $N$-body Milky Way-like cosmological simulations with embedded disk potentials along with the full phase-space information for Milky Way satellites from {\it Gaia} EDR3, we constrain the dark matter halo mass of the Milky Way and find a preferred mass range of $\sim1$--$1.2 \times 10^{12}~\msun$. A more complete summary of our main results are as follows:

\renewcommand{\labelenumii}{(\roman{enumii})}
\begin{enumerate}
    \item As illustrated in Fig.~\ref{fig:lvterr_vels}, when limiting the observed sample of Milky Way satellites to those systems with well-measured kinematics, we find that the observed distribution of satellite velocities ($V_{\rm rad}$, $V_{\rm tan}$, \emph{and} $V_{\rm tot}$) are consistent with a host halo mass of $\sim 1-1.2 \times 10^{12}~\msun$.

    \item Across all samples probed, the distribution of satellite velocities inferred from {\it Gaia} observations of the Milky Way satellites are inconsistent, at the $3\sigma$ confidence level, with that of subhalos populating host halos with masses $<10^{12}~\msun$ or $>1.2 \times 10^{12}~\msun$. Our use of the MWU test allows us to quantify the inconsistency or "anti–goodness of fit".

    \item Excluding systems associated with the LMC does not significantly change our results, with the observed kinematics of the Milky Way satellites favoring a host halo mass of $\sim 1-1.2 \times 10^{12}~\msun$ when compared to distance-matched subhalo populations in phELVIS.

    \item In the inner halo ($D_{\rm MW} < 100$~kpc), we find a correlation between host mass and the eccentricity of satellite orbits (as predicted by {\sc galpy}), such that at a given Galactocentric distance increasingly circular orbits are found in higher-mass hosts. This is likely a consequence of subhalo destruction preferentially removing satellites on more radial orbits in more massive hosts. 

    \item The distribution of infall times inferred from {\it Gaia} phase-space measures \citep{fham19} are systematically skewed towards early cosmic times (i.e.~early accretion) relative to that of distance-matched subhalos drawn from the phELVIS simulation suite. 

    \item The distribution of pericentric distances for subhalos in phELVIS show little dependence on host mass, in contrast to the expectations from {\sc galpy} that favor smaller pericentric distances for satellites in more massive host halos. 

    \item Looking towards the discovery of future Milky Way satellites by next-generation observational facilities, we show that the observed distribution of Galactocentric total velocity and Galactocentric distance stand to be good metrics to test the preferred host mass range for the Milky Way.
\end{enumerate}

\section*{Acknowledgements}

We thank Tyler Kelley and Marcel Pawlowski for helpful discussions regarding this project.
This work was supported in part by NSF grants AST-1815475 and AST-1518257.  
MKRW acknowledges support from the National Science Foundation Graduate Research Fellowship and MPS–Ascend Postdoctoral Research Fellowship. This material is based upon work supported by the National Science Foundation Graduate Research Fellowship Program under Grant No.~DGE-1321846. This material is based upon work supported by the National Science Foundation MPS-Ascend Postdoctoral Research Fellowship under Grant No. ~AST-2138144.
MBK acknowledges support from NSF CAREER award AST-1752913, NSF grant AST-1910346, NASA grant NNX17AG29G, and HST-AR-15006, HST-AR-15809, HST-GO-15658, HST-GO-15901, HST-GO-15902, HST-AR-16159, and HST-GO-16226 from the Space Telescope Science Institute, which is operated by AURA, Inc., under NASA contract NAS5-26555.
LVS acknowledges support from the NASA ATP 80NSSC20K0566 and NSF CAREER 1945310 grants. 
DCB thanks the LSSTC Data Science Fellowship Program, which is funded by LSSTC, NSF Cybertraining Grant $\#$1829740, the Brinson Foundation, and the Moore Foundation; participation in the program has greatly benefited this work.

This research has made use of NASA’s Astrophysics Data System Bibliographic Services. This research also utilized \texttt{astropy}, a community-developed core Python package for Astronomy \citep{astropy13, astropy18}. Additionally, the Python packages NumPy \citep{numpy}, iPython \citep{ipython}, SciPy \citep{scipy}, matplotlib \citep{hunter07}, and scikit-learn \citep{pedregosa12} were utilized for our data analysis and presentation.

This work has made use of data from the European Space Agency (ESA) mission {\it Gaia} (\url{https://www.cosmos.esa.int/gaia}), processed by the {\it Gaia} Data Processing and Analysis Consortium (DPAC, \url{https://www.cosmos.esa.int/web/gaia/dpac/consortium}). Funding for the DPAC has been provided by national institutions, in particular the institutions participating in the {\it Gaia} Multilateral Agreement.

\section*{Data Availability}
The data used within this work is available from the corresponding author upon request.

\bibliographystyle{mnras}
\bibliography{MKRW_Sizing} 


\clearpage 

\begin{table*}
    \begin{center}
    \begin{tabular}[h!]{@{}ccccccccc@{}}
        \hline
        \hline
        Galaxy & Confirmed? & $\mu_{\alpha}$ cos $\delta$ & $\mu_{\delta}$ & $D$$_{\rm MW}$ & $V$$_{\rm rad}$ & $V$$_{\rm tan}$ & $V$$_{\rm 3D}$ & Notes \\
         &  & (mas / yr) & (mas / yr) & (kpc) & (km / s) & (km / s) & (km / s) &  \\
        \hline
        Antlia II & Y & -0.09$\pm$0.01 & 0.12$\pm$0.01 & 133.0$\pm$6.0 & 70.4$\pm$0.5 & 125.0$\pm$6.1 & 143.5$\pm$5.3 & 1 \\
        Aquarius II & Y & -0.17$\pm$0.1 & -0.43$\pm$0.08 & 105.3$\pm$3.3 & 30.4$\pm$7.2 & 157.0$\pm$51.6 & 159.9$\pm$50.7 & 3 \\
        Boötes I & Y & -0.39$\pm$0.01 & -1.06$\pm$0.01 & 63.6$\pm$2.0 & 91.2$\pm$2.1 & 156.0$\pm$3.0 & 180.7$\pm$2.8 & 1 \\
        Boötes II & Y & -2.33${^{+0.09}_{-0.08}}$ & -0.41$\pm$0.06 & 39.8$\pm$1.0 & -54.3$\pm$3.9 & 319.0${^{+17.9}_{-16.0}}$ & 323.6${^{+17.6}_{-15.8}}$ & 1 \\
        Canes Venatici I & Y & -0.11$\pm$0.02 & -0.12$\pm$0.02 & 210.8$\pm$6.0 & 78.2$\pm$0.5 & 69.0$\pm$22.5 & 104.3$\pm$14.9 &   \\
        Canes Venatici II & Y & -0.15$\pm$0.07 & -0.27$\pm$0.06 & 160.6$\pm$4.0 & -96.7$\pm$0.2 & 31.0$\pm$62.9 & 101.5$\pm$19.2 &   \\
        Carina & Y & 0.53$\pm$0.01 & 0.12$\pm$0.01 & 107.6$\pm$5.0 & 8.5$\pm$0.3 & 187.0$\pm$7.6 & 187.2$\pm$7.6 & 1 \\
        Carina II & Y & 1.88$\pm$0.01 & 0.13$\pm$0.02 & 37.1$\pm$0.6 & 219.5$\pm$1.9 & 268.0$\pm$3.8 & 346.4$\pm$3.2 & 1,2 \\
        Carina III & Y & 3.12$\pm$0.05 & 1.54${^{+0.06}_{-0.07}}$ & 29.0$\pm$0.6 & 58.7$\pm$4.7 & 395.0${^{+12.4}_{-12.5}}$ & 399.3${^{+12.3}_{-12.4}}$ & 1,2 \\
        Columba I & P & 0.19$\pm$0.06 & -0.36$\pm$0.06 & 187.6$\pm$10.0 & -22.8$\pm$6.8 & 205.0$\pm$52.4 & 206.3$\pm$52.1 &   \\
        Coma Berenices & Y & 0.41$\pm$0.02 & -1.71$\pm$0.02 & 43.2$\pm$1.5 & 31.9$\pm$0.7 & 264.0$\pm$4.2 & 265.9$\pm$4.2 & 1 \\
        Crater II & Y & -0.07$\pm$0.02 & -0.11$\pm$0.01 & 116.4$\pm$1.1 & -76.0$\pm$-0.1 & 102.0$\pm$9.4 & 127.2$\pm$7.5 & 1 \\
        Draco & Y & 0.042$\pm$0.005 & -0.19$\pm$0.01 & 82.0$\pm$6.0 & -103.0$\pm$0.4 & 156.0$\pm$1.4 & 186.9$\pm$1.2 & 1 \\
        Draco II & P & 1.08$\pm$0.07 & 0.91$\pm$0.08 & 23.9$\pm$3.8 & -156.6$\pm$1.6 & 299.0$\pm$15.0 & 337.5$\pm$13.3 & 1 \\
        Fornax & Y & 0.382$\pm$0.001 & -0.359$\pm$0.002 & 141.1$\pm$3.0 & -34.5$\pm$0.2 & 142.0$\pm$0.0 & 146.1${^{+5.3}_{-0.0}}$ & 1 \\
        Grus I & P & 0.07$\pm$0.05 & -0.29${^{+0.06}_{-0.07}}$ & 116.2$\pm$11.5 & -187.3$\pm$4.3 & 71.0${^{+38.0}_{-41.3}}$ & 200.3${^{+14.1}_{-15.2}}$ &   \\
        Grus II & P & 0.38$\pm$0.03 & -1.46$\pm$0.04 & 48.4$\pm$5.0 & -124.8$\pm$1.7 & 139.0$\pm$10.2 & 186.8$\pm$7.7 & 1 \\
        Hercules & Y & -0.03$\pm$0.04 & -0.36$\pm$0.03 & 126.3$\pm$6.0 & 141.1$\pm$1.3 & 146.0$\pm$24.4 & 203.0$\pm$17.6 & 1 \\
        Horologium I & Y & 0.82$\pm$0.03 & -0.61$\pm$0.03 & 87.3$\pm$12.0 & -28.4$\pm$4.3 & 193.0$\pm$15.9 & 195.1$\pm$15.8 & 1,2 \\
        Horologium II & P & 0.76${^{+0.2}_{-0.29}}$ & -0.41${^{+0.23}_{-0.21}}$ & 79.1$\pm$7.5 & 29.1$\pm$25.0 & 128.0${^{+111.9}_{-157.7}}$ & 131.3${^{+109.3}_{-153.8}}$ &   \\
        Hydra II & P & -0.34$\pm$0.1 & -0.09${^{+0.08}_{-0.09}}$ & 148.1$\pm$7.5 & 136.6$\pm$0.7 & 97.0${^{+71.8}_{-72.6}}$ & 167.5${^{+41.6}_{-42.0}}$ &   \\
        Hydrus I & Y & 3.79$\pm$0.01 & -1.5$\pm$0.01 & 25.7$\pm$0.5 & -40.1$\pm$1.4 & 363.0$\pm$5.0 & 365.2$\pm$4.9 & 1,2 \\
        Leo I & Y & -0.05$\pm$0.01 & -0.11$\pm$0.01 & 257.9$\pm$15.5 & 174.7$\pm$0.1 & 75.0$\pm$12.1 & 190.1$\pm$4.8 & 1 \\
        Leo II & Y & -0.14$\pm$0.02 & -0.12$\pm$0.02 & 235.6$\pm$14.0 & 26.5$\pm$0.3 & 103.0$\pm$22.6 & 106.3$\pm$21.9 & 1 \\
        Leo IV & Y & -0.08$\pm$0.09 & -0.21$\pm$0.08 & 154.6$\pm$5.0 & 8.6$\pm$0.2 & 52.0$\pm$63.8 & 52.7$\pm$62.9 &   \\
        Leo V & Y & -0.06$\pm$0.09 & -0.25${^{+0.09}_{-0.08}}$ & 169.8$\pm$4.0 & 55.8$\pm$1.6 & 69.0${^{+83.5}_{-78.6}}$ & 88.8${^{+64.9}_{-61.1}}$ &   \\
        Phoenix II & Y & 0.48$\pm$0.04 & -1.17$\pm$0.05 & 81.3$\pm$4.0 & -38.1$\pm$5.9 & 271.0$\pm$20.7 & 273.7$\pm$20.6 & 1,3 \\
        Pisces II & Y & 0.11$\pm$0.11 & -0.24${^{+0.12}_{-0.11}}$ & 182.1$\pm$15.0 & -75.7$\pm$8.7 & 47.0${^{+102.1}_{-93.7}}$ & 89.1${^{+54.3}_{-49.9}}$ &   \\
        Reticulum II & Y & 2.39$\pm$0.01 & -1.36$\pm$0.02 & 33.0$\pm$1.4 & -92.2$\pm$1.7 & 214.0$\pm$2.2 & 233.0$\pm$2.1 & 1,3 \\
        Reticulum III & P & 0.36$\pm$0.14 & 0.05${^{+0.19}_{-0.25}}$ & 92.0$\pm$13.0 & 113.4$\pm$17.9 & 78.0${^{+83.1}_{-109.1}}$ & 137.6${^{+49.4}_{-63.6}}$ &   \\
        Sagittarius II & Y & -0.77$\pm$0.03 & -0.89$\pm$0.02 & 63.0$\pm$2.3 & -115.7$\pm$1.8 & 239.0$\pm$10.4 & 265.5$\pm$9.4 & 1 \\
        Sculptor & Y & 0.099$\pm$0.002 & -0.16$\pm$0.002 & 86.1$\pm$5.0 & 76.3$\pm$0.3 & 163.0$\pm$0.0 & 180.0$\pm$0.1 & 1,3 \\
        Segue 1 & Y & -2.21$\pm$0.06 & -3.34$\pm$0.05 & 28.0$\pm$1.9 & 136.9$\pm$0.0 & 240.0$\pm$6.5 & 276.3$\pm$5.7 & 1,3 \\
        Segue 2 & Y & 1.47$\pm$0.04 & -0.31$\pm$0.04 & 43.1$\pm$3.0 & 59.9$\pm$5.0 & 134.0$\pm$7.2 & 146.8$\pm$6.9 & 1 \\
        Sextans I & Y & -0.41$\pm$0.01 & 0.04$\pm$0.01 & 98.1$\pm$3.0 & 88.1$\pm$0.1 & 220.0$\pm$4.0 & 237.0$\pm$3.7 & 1 \\
        Triangulum II & P & 0.56$\pm$0.05 & 0.07$\pm$0.06 & 34.8$\pm$1.6 & -265.6$\pm$3.2 & 159.0$\pm$8.9 & 309.5$\pm$5.3 & 1 \\
        Tucana II & Y & 0.9$\pm$0.02 & -1.26$\pm$0.02 & 54.2$\pm$7.9 & -182.0$\pm$4.6 & 210.0$\pm$8.1 & 277.9$\pm$6.9 & 1 \\
        Tucana III & P & -0.08$\pm$0.01 & -1.62$\pm$0.02 & 23.0$\pm$1.9 & -223.4$\pm$2.6 & 126.0$\pm$2.0 & 256.5$\pm$2.5 & 1,3 \\
        Tucana IV & Y & 0.54$\pm$0.06 & -1.67$\pm$0.07 & 45.4$\pm$3.9 & -90.6$\pm$5.3 & 197.0$\pm$16.0 & 216.9$\pm$14.7 & 1 \\
        Tucana V & P & -0.14${^{+0.06}_{-0.05}}$ & -1.15${^{+0.08}_{-0.06}}$ & 51.8$\pm$8.9 & -157.8$\pm$5.9 & 181.0${^{+30.9}_{-24.9}}$ & 240.1${^{+23.6}_{-19.1}}$ & 1 \\
        Ursa Major I & Y & -0.39$\pm$0.03 & -0.63$\pm$0.03 & 102.1$\pm$5.8 & -0.8$\pm$0.7 & 126.0$\pm$20.8 & 126.0$\pm$20.8 & 1 \\
        Ursa Major II & Y & 1.72$\pm$0.02 & -1.89$\pm$0.03 & 41.0$\pm$1.9 & -64.5$\pm$1.6 & 262.0$\pm$6.5 & 269.8$\pm$6.4 & 1 \\
        Ursa Minor & Y & -0.124$\pm$0.004 & 0.078$\pm$0.04 & 77.9$\pm$4.0 & -83.4$\pm$1.4 & 148.0$\pm$0.0 & 169.9$\pm$0.7 & 1 \\
        Willman 1 & Y & 0.21$\pm$0.06 & -1.08$\pm$0.09 & 49.7$\pm$9.9 & 16.7$\pm$1.9 & 120.0$\pm$17.5 & 121.2$\pm$17.3 & 1 \\
        \hline
    \end{tabular}
    \caption{Properties of the MW Satellite Galaxies used in this work. Column (1) Status of whether the system is a spectroscopically confirmed galaxy or not (i.e. Y $=$ confirmed galaxy and P $=$ not confirmed but probably a galaxy). Columns (2) $\&$ (3) Proper motions derived by \citet{mcconnedr3} from {\it Gaia} EDR3 in mas~yr$^{-1}$. Column (4) Galactocentric distance with errors in kpc. Columns (5) - (7) Radial, tangential and total velocities, respectively, in the Galactocentric frame of reference, all in km~s$^{-1}$. Galactocentric distance and radial velocity were converted from the heliocentric frame of reference using \texttt{astropy} and quantities from MCV20a, MCV20b and the heliocentric distances referenced in \S\ref{subsec:gaia}. Galactocentric tangential velocity was converted from the Galactocentric tangential velocity components provided in MCV20a. Total velocity was then calculated from its two components. Column (8) indicates the various subgroups a galaxy belongs to which are used throughout this work. 1 indicates belonging to the group of systems with proportionally low tangential velocity errors. 2 indicates the system is a long term satellite of the LMC as determined by \citet{patel20}. 3 indicates the system is a short term satellite or recent interactor with the LMC as determined by \citet{patel20}.}
    \label{table:1a}
    \end{center}
\end{table*}

\newpage

\begin{table*}
    \begin{center}
    \begin{tabular}[h!]{@{}cccccccc@{}}
        \hline
        Galaxy & $D$$_{\rm peri,l}$ & $D$$_{\rm peri,i}$ & $D$$_{\rm peri,h}$ & $e_{\rm l}$ & $e_{\rm i}$ & $e_{\rm h}$ & $t$$_{\rm infall}$ \\
         & (kpc) & (kpc) & (kpc) &  &  &  & (Gyr) \\
        \hline
        Antlia II & 69.8$\pm$13.1 & 54.9$\pm$10.8 & 43.4$\pm$8.3 & 0.4$\pm$-0.0 & 0.5$\pm$-0.0 & 0.5$\pm$-0.0 & -- \\
        Aquarius II & 92.7$\pm$-45.2 & 72.5$\pm$-35.8 & 54.6$\pm$-25.5 & 0.2$\pm$0.3 & 0.2$\pm$0.3 & 0.3$\pm$0.3 & 1.6${^{+5.4}_{-3.5}}$ \\
        Boötes I & 43.9$\pm$3.1 & 37.4$\pm$3.2 & 30.8$\pm$2.9 & 0.4$\pm$0.0 & 0.4$\pm$-0.0 & 0.4$\pm$-0.0 & 10.7${^{+0.6}_{-1.9}}$ \\
        Boötes II & 39.1$\pm$1.0 & 39.0$\pm$1.0 & 38.8$\pm$1.1 & 0.9$\pm$-0.0 & 0.7$\pm$-0.0 & 0.4$\pm$-0.0 & 1.1$\pm$0.6 \\
        Canes Venatici I & 47.4$\pm$33.4 & 37.8$\pm$25.7 & 30.7$\pm$19.8 & 0.7$\pm$-0.1 & 0.7$\pm$-0.1 & 0.8$\pm$-0.1 & 9.4${^{+0.9}_{-2.3}}$ \\
        Canes Venatici II & 9.8$\pm$32.0 & 8.3$\pm$25.5 & 7.2$\pm$20.4 & 0.9$\pm$-0.2 & 0.9$\pm$-0.2 & 0.9$\pm$-0.2 & 9.0${^{+1.0}_{-2.8}}$ \\
        Carina & 107.3$\pm$5.1 & 106.5$\pm$5.7 & 83.6$\pm$25.3 & 0.3$\pm$0.2 & 0.1$\pm$0.1 & 0.1$\pm$-0.1 & 9.9${^{+0.6}_{-2.7}}$ \\
        Carina II & 37.1$\pm$0.6 & 27.0$\pm$0.7 & 25.9$\pm$0.7 & 0.9$\pm$0.0 & 0.8$\pm$0.0 & 0.7$\pm$0.0 & 7.9${^{+2.5}_{-2.4}}$ \\
        Carina III & 28.9$\pm$0.6 & 28.5$\pm$0.6 & 28.4$\pm$0.6 & 1.0$\pm$0.0 & 0.8$\pm$0.1 & 0.6$\pm$0.1 & 7.6${^{+2.4}_{-2.7}}$ \\
        Columba I & 185.7$\pm$10.7 & 185.0$\pm$10.8 & 182.8$\pm$10.7 & 0.7$\pm$-0.0 & 0.5$\pm$-0.1 & 0.2$\pm$-0.1 & -- \\
        Coma Berenices & 42.5$\pm$1.6 & 42.3$\pm$1.6 & 41.4$\pm$1.9 & 0.5$\pm$0.1 & 0.3$\pm$0.1 & 0.1$\pm$0.0 & 10.2${^{+2.6}_{-3.3}}$ \\
        Crater II & 39.9$\pm$8.4 & 31.9$\pm$6.5 & 25.8$\pm$5.0 & 0.6$\pm$-0.1 & 0.6$\pm$-0.1 & 0.7$\pm$-0.0 & 7.8${^{+2.7}_{-3.0}}$ \\
        Draco & 49.0$\pm$5.9 & 41.2$\pm$5.3 & 33.8$\pm$4.3 & 0.5$\pm$0.0 & 0.5$\pm$-0.0 & 0.5$\pm$-0.0 & 10.4${^{+2.4}_{-3.1}}$ \\
        Draco II & 19.9$\pm$4.3 & 19.4$\pm$4.4 & 18.7$\pm$4.6 & 0.8$\pm$0.1 & 0.6$\pm$0.1 & 0.5$\pm$0.1 & 10.2${^{+1.8}_{-2.4}}$ \\
        Fornax & 98.7$\pm$12.6 & 73.4$\pm$10.1 & 55.9$\pm$7.2 & 0.2$\pm$-0.0 & 0.3$\pm$-0.0 & 0.4$\pm$-0.0 & 10.7${^{+0.8}_{-3.1}}$ \\
        Grus I & 17.1$\pm$9.2 & 14.7$\pm$7.8 & 12.7$\pm$6.5 & 0.9$\pm$-0.0 & 0.9$\pm$-0.0 & 0.9$\pm$-0.0 & 1.1${^{+1.0}_{-0.9}}$ \\
        Grus II & 26.0$\pm$9.4 & 22.4$\pm$9.0 & 18.9$\pm$7.8 & 0.5$\pm$0.0 & 0.5$\pm$-0.1 & 0.5$\pm$-0.1 & -- \\
        Hercules & 70.2$\pm$62.3 & 61.4$\pm$13.2 & 52.0$\pm$12.5 & 0.7$\pm$-0.1 & 0.6$\pm$0.0 & 0.6$\pm$-0.0 & 6.6${^{+2.3}_{-0.7}}$ \\
        Horologium I & 86.1$\pm$12.7 & 85.6$\pm$13.1 & 83.3$\pm$15.2 & 0.6$\pm$0.3 & 0.3$\pm$0.4 & 0.1$\pm$0.3 & 8.8${^{+1.8}_{-2.0}}$ \\
        Horologium II & 49.3$\pm$37.2 & 37.4$\pm$44.9 & 29.3$\pm$50.4 & 0.3$\pm$0.5 & 0.4$\pm$0.1 & 0.5$\pm$-0.2 & -- \\
        Hydra II & 62.6$\pm$-7.5 & 52.8$\pm$-6.8 & 43.8$\pm$-5.7 & 0.7$\pm$0.0 & 0.7$\pm$0.0 & 0.7$\pm$0.0 & 9.4${^{+1.7}_{-1.8}}$ \\
        Hydrus I & 25.5$\pm$0.5 & 25.5$\pm$0.5 & 25.5$\pm$0.5 & 0.9$\pm$0.0 & 0.7$\pm$0.0 & 0.5$\pm$0.0 & 10.7${^{+1.3}_{-1.4}}$ \\
        Leo I & 257.7$\pm$15.5 & 257.7$\pm$15.5 & 35.6$\pm$6.4 & 0.7$\pm$0.0 & 0.5$\pm$0.0 & 0.9$\pm$-0.0 & 10.5${^{+1.5}_{-2.4}}$ \\
        Leo II & 120.3$\pm$24.0 & 88.1$\pm$16.4 & 67.2$\pm$11.6 & 0.3$\pm$-0.1 & 0.5$\pm$-0.0 & 0.6$\pm$-0.0 & 2.3${^{+0.6}_{-0.5}}$ \\
        Leo IV & 20.2$\pm$107.7 & 16.5$\pm$72.9 & 13.7$\pm$52.7 & 0.8$\pm$-0.7 & 0.8$\pm$-0.5 & 0.8$\pm$-0.4 & 7.8${^{+3.3}_{-2.0}}$ \\
        Leo V & 32.5$\pm$89.2 & 26.1$\pm$70.1 & 21.3$\pm$52.8 & 0.7$\pm$-0.4 & 0.7$\pm$-0.4 & 0.8$\pm$-0.3 & 10.4$\pm$1.4 \\
        Phoenix II & 80.3$\pm$4.3 & 80.1$\pm$4.4 & 79.6$\pm$4.6 & 0.9$\pm$0.0 & 0.6$\pm$0.1 & 0.4$\pm$0.1 & -- \\
        Pisces II & 28.7$\pm$108.7 & 23.3$\pm$88.5 & 19.2$\pm$68.0 & 0.8$\pm$-0.4 & 0.8$\pm$-0.4 & 0.8$\pm$-0.4 & -- \\
        Reticulum II & 28.7$\pm$2.3 & 27.3$\pm$2.7 & 24.8$\pm$3.3 & 0.5$\pm$0.1 & 0.3$\pm$0.0 & 0.3$\pm$0.0 & 8.3$\pm$1.8 \\
        Reticulum III & 17.2$\pm$87.7 & 14.4$\pm$58.5 & 12.1$\pm$52.9 & 0.8$\pm$-0.1 & 0.8$\pm$-0.2 & 0.8$\pm$-0.3 & -- \\
        Sagittarius II & 51.1$\pm$2.1 & 48.0$\pm$2.0 & 43.1$\pm$1.8 & 0.6$\pm$0.0 & 0.5$\pm$-0.0 & 0.4$\pm$-0.0 & -- \\
        Sculptor & 58.9$\pm$1.6 & 48.4$\pm$1.1 & 38.8$\pm$0.8 & 0.4$\pm$0.0 & 0.4$\pm$0.0 & 0.4$\pm$0.0 & 10.6${^{+1.6}_{-1.9}}$ \\
        Segue 1 & 21.4$\pm$3.1 & 20.2$\pm$3.5 & 18.5$\pm$3.8 & 0.6$\pm$0.1 & 0.5$\pm$0.1 & 0.4$\pm$0.0 & 9.9${^{+1.7}_{-2.9}}$ \\
        Segue 2 & 22.1$\pm$8.3 & 17.9$\pm$6.8 & 14.7$\pm$5.4 & 0.4$\pm$-0.1 & 0.4$\pm$-0.1 & 0.5$\pm$-0.1 & 10.8${^{+1.3}_{-1.4}}$ \\
        Sextans I & 97.9$\pm$3.0 & 82.5$\pm$3.9 & 75.4$\pm$4.5 & 0.6$\pm$0.1 & 0.4$\pm$0.0 & 0.3$\pm$0.0 & 10.8${^{+1.6}_{-1.9}}$ \\
        Triangulum II & 11.7$\pm$1.3 & 10.7$\pm$1.2 & 9.7$\pm$1.1 & 0.9$\pm$-0.0 & 0.8$\pm$-0.0 & 0.8$\pm$-0.0 & 8.4${^{+2.7}_{-0.9}}$ \\
        Tucana II & 38.6$\pm$11.9 & 36.6$\pm$12.6 & 33.8$\pm$13.4 & 0.9$\pm$0.1 & 0.7$\pm$0.2 & 0.6$\pm$0.1 & 9.5${^{+1.5}_{-2.1}}$ \\
        Tucana III & 3.0$\pm$0.6 & 2.8$\pm$0.4 & 2.5$\pm$0.5 & 0.9$\pm$-0.0 & 0.9$\pm$-0.0 & 0.9$\pm$-0.0 & -- \\
        Tucana IV & 36.1$\pm$6.3 & 32.5$\pm$7.1 & 27.6$\pm$7.0 & 0.4$\pm$0.1 & 0.3$\pm$-0.0 & 0.3$\pm$-0.0 & -- \\
        Tucana V & 27.9$\pm$9.4 & 24.7$\pm$8.7 & 21.4$\pm$7.7 & 0.6$\pm$0.0 & 0.6$\pm$-0.0 & 0.6$\pm$-0.0 & -- \\
        Ursa Major I & 70.6$\pm$3.7 & 51.4$\pm$2.6 & 39.3$\pm$1.9 & 0.2$\pm$0.0 & 0.3$\pm$0.0 & 0.4$\pm$0.0 & 9.5${^{+2.4}_{-2.8}}$ \\
        Ursa Major II & 39.3$\pm$2.2 & 38.9$\pm$2.4 & 38.1$\pm$2.7 & 0.7$\pm$0.1 & 0.5$\pm$0.1 & 0.3$\pm$0.1 & 1.5${^{+5.1}_{-1.6}}$ \\
        Ursa Minor & 44.8$\pm$8.1 & 36.4$\pm$7.0 & 29.5$\pm$5.6 & 0.4$\pm$-0.0 & 0.4$\pm$-0.0 & 0.5$\pm$-0.0 & 10.7${^{+1.4}_{-2.3}}$ \\
        Willman 1 & 24.6$\pm$21.4 & 19.4$\pm$15.2 & 15.7$\pm$11.2 & 0.3$\pm$-0.2 & 0.4$\pm$-0.2 & 0.5$\pm$-0.1 & 10.7${^{+1.7}_{-2.0}}$ \\
        \hline
    \end{tabular}
    \caption{Orbital and Infall Properties of the MW Satellite Galaxies used in this work. Columns (1) - (3) Pericentric passage distances, in kpc, for each system in a low, intermediate and high MW potential, with an NFW profile and concentration of $15.3$, via {\sc galpy}. The corresponding MW dark matter halo masses used for the $3$ potentials are $0.835, 1.120, 1.675 \times 10^{12}~\msun$, respectively. These are the average masses of the hosts in each of our fiducial phELVIS host sets. Columns (4) - (6) Orbital eccentricities for each system in the three {\sc galpy} MW potentials, with smaller eccentricities corresponding to more circular orbits and larger eccentricities corresponding to more radial/plunging orbits. Column (7) Satellite infall times, in Gyr, as derived by \citet{fham19}.}    
    \label{table:1b}
    \end{center}
\end{table*}

\newpage

\begin{table*}
\label{table:pvales}
    \begin{center}
    \begin{tabular}[h!]{@{}ccccc@{}}
        \hline
        & D$_{\rm MW}$ & V$_{\rm Rad}$ & V$_{\rm Tan}$ & V$_{\rm 3D}$ \\ [0.5ex] 
        \hline\hline 
        \multicolumn{5}{c}{\textbf{All Satellites (44) to Unmatched Subhalos}}\\
        Low Mass Hosts & $<0.001$ & $<0.001$ & $<0.001$ & $<0.001$ \\
        Intermediate Mass Hosts & $<0.001$ & 0.003 & $<0.001$ & $<0.001$ \\
        High Mass Hosts & $<0.001$ & \cellcolor[HTML]{F2D25B} 0.367 & $<0.001$ & $<0.001$ \\
        \hline\hline 
        \multicolumn{5}{c}{\textbf{Satellites with Proportionally Low}}\\
        \multicolumn{5}{c}{\textbf{Tangential Velocity Errors (LTVE) (34) to Unmatched Subhalos}}\\
        Low Mass Hosts & $<0.001$ & $<0.001$ & $<0.001$ & $<0.001$ \\
        Intermediate Mass Hosts & $<0.001$ & 0.001 & $<0.001$ & $<0.001$ \\
        High Mass Hosts & $<0.001$ & \cellcolor[HTML]{F2D25B} 0.384 & $<0.001$ & $<0.001$ \\
        \hline\hline 
        \multicolumn{5}{c}{\textbf{All Satellites (44)}}\\
        Low Mass Hosts & \cellcolor[HTML]{F2D25B} 0.836 & $<0.001$ & \cellcolor[HTML]{F2D25B} 0.142 & 0.006 \\
        Intermediate Mass Hosts & \cellcolor[HTML]{F2D25B} 0.822 & \cellcolor[HTML]{F2D25B} 0.735 & 0.005 & 0.013 \\
        High Mass Hosts & \cellcolor[HTML]{F2D25B} 0.796 & $<0.001$ & $<0.001$ & $<0.001$ \\
        \hline\hline 
        \multicolumn{5}{c}{\textbf{Satellites with Proportionally Low}}\\
        \multicolumn{5}{c}{\textbf{Tangential Velocity Errors (LTVE) (34)}}\\
        Low Mass Hosts & \cellcolor[HTML]{F2D25B} 0.767 & $<0.001$ & 0.003 & $<0.001$ \\
        Intermediate Mass Hosts & \cellcolor[HTML]{F2D25B} 0.744 & \cellcolor[HTML]{F2D25B} 0.896 & \cellcolor[HTML]{F2D25B} 0.098 & \cellcolor[HTML]{F2D25B} 0.141 \\
        High Mass Hosts & \cellcolor[HTML]{F2D25B} 0.702 & $<0.001$ & $<0.001$ & $<0.001$ \\
        \hline\hline 
        \multicolumn{5}{c}{\textbf{LTVE Satellites Excluding}}\\
        \multicolumn{5}{c}{\textbf{Satellites Associated with the LMC (25)}}\\
        Low Mass Hosts & \cellcolor[HTML]{F2D25B} 0.823 & 0.003 & 0.019 & 0.003 \\
        Intermediate Mass Hosts & \cellcolor[HTML]{F2D25B} 0.759 & \cellcolor[HTML]{F2D25B} 0.722 & \cellcolor[HTML]{F2D25B} 0.081 &\cellcolor[HTML]{F2D25B} 0.057\\
        High Mass Hosts & \cellcolor[HTML]{F2D25B} 0.778 & 0.001 & $<0.001$ & $<0.001$ \\
        \hline\hline 
        \multicolumn{5}{c}{\textbf{LTVE Satellites with \boldmath$D_{\rm MW} < 100$ kpc (25)}}\\
        Low Mass Hosts & \cellcolor[HTML]{F2D25B} 0.644 & $<0.001$ & 0.012 & $<0.001$ \\
        Intermediate Mass Hosts & \cellcolor[HTML]{F2D25B} 0.641 & \cellcolor[HTML]{F2D25B} 0.061 & 0.005 & 0.044 \\
        High Mass Hosts & \cellcolor[HTML]{F2D25B} 0.565 & 0.032 & $<0.001$ & $<0.001$ \\
        \hline\hline 
        \multicolumn{5}{c}{\textbf{LTVE Satellites Excluding Satellites Associated}}\\
        \multicolumn{5}{c}{\textbf{ with the LMC and using limit of \boldmath$V_{\rm peak} > 10~{\rm km}~{\rm s}^{-1}$  (25)}}\\
        Low Mass Hosts & \cellcolor[HTML]{F2D25B} 0.769 & \cellcolor[HTML]{F2D25B} 0.060 & 0.015 & 0.003 \\
        Intermediate Mass Hosts & \cellcolor[HTML]{F2D25B} 0.798 & \cellcolor[HTML]{F2D25B} 0.055 & 0.049 & 0.008 \\
        High Mass Hosts & \cellcolor[HTML]{F2D25B} 0.730 & $<0.001$ & $<0.001$ & $<0.001$ \\
        \hline
    \end{tabular}
    \caption{Mann-Whitney $U$ Test $p$-values for the observation to distance-matched subhalo comparisons. The reported $p$-values are the harmonic means of $500$ MWU Tests conducted on sets of $10N$ distance-matched subhalo properties and $10N$ MW satellite properties drawn from the properties' errors. These $p$-values are two-sided, i.e. $p_{\rm max} = 1$. Our statistical significance level is set at $p = 0.05$. Values below this level reject the null hypothesis that the two compared distributions are drawn from the same parent population. Values above our chosen significance level signify that the null hypothesis cannot be rejected and are highlighted in yellow. Values are reported for our fiducial phELVIS host sets for all the various groups of satellites discussed throughout the paper.}
    \label{table:2}
    \end{center}
\end{table*}

\clearpage

\bsp	
\label{lastpage}
\end{document}